\documentclass[final,5p,times,twocolumn]{elsarticle} %1p / 5p
\DeclareMathAlphabet{\mathpzc}{OT1}{pzc}{m}{it}
\usepackage{lineno}
\usepackage{setspace}
\usepackage{graphicx}
\usepackage{amssymb}
\usepackage{amsmath}
\usepackage{amsthm}
\usepackage[colorlinks=true]{hyperref}
\usepackage[all]{hypcap}
\usepackage{mathrsfs}

\usepackage[percent]{overpic}
\usepackage{fixltx2e}
\usepackage{epsfig}
\usepackage{verbatim}
\usepackage{multirow}
\usepackage{float}
\usepackage{array}
\usepackage{alltt}
\usepackage{relsize}
\usepackage{upgreek}
\usepackage{arydshln}

\newcommand{\car}{\text{nat}}

\newcommand{\R}{\mathcal{R}}
\newcommand{\pair}{\mathcal{P}}
\newcommand{\Z}{\mathcal{C}}
\newcommand{\rel}{r}

\journal{Nuclear Instruments and Methods A}
\begin{document}
\begin{frontmatter}
%\title{Neutron background estimation methods}

\title{Machine learning based event classification for the energy-differential measurement of the $^\car$C(\textit{n,p}) and $^\car$C(\textit{n,d}) reactions}

%Difference between the neutron sensitivity and the neutron background

\author[a]{P.~\v{Z}ugec\corref{cor1}}\ead{pzugec@phy.hr}
\author[b,c]{M.~Barbagallo}
\author[d]{J.~Andrzejewski}
\author[d]{J.~Perkowski}
\author[b]{N.~Colonna}
\author[a]{D.~Bosnar}
\author[d]{A.~Gawlik}
\author[c,e]{M.~Sabat\'{e}-Gilarte}
\author[c,f]{M.~Bacak}
\author[c]{F.~Mingrone}
\author[c]{E.~Chiaveri}

\address[a]{Department of Physics, Faculty of Science, University of Zagreb, Croatia}
\address[b]{Istituto Nazionale di Fisica Nucleare, Sezione di Bari, Italy}
\address[c]{European Organization for Nuclear Research (CERN), Geneva, Switzerland}
\address[d]{Uniwersytet \L\'{o}dzki, Lodz, Poland}
\address[e]{Universidad de Sevilla, Spain}
\address[f]{Technische Universit\"{a}t Wien, Vienna, Austria}

%\address[g]{Istituto Nazionale di Fisica Nucleare, Sezione di Bologna, Italy}

\author{\linebreak The n\_TOF Collaboration\fnref{fn1}} 
\cortext[cor1]{Corresponding author. Tel.: +385 1 4605552}
\fntext[fn1]{www.cern.ch/ntof}

\begin{abstract}

The paper explores the feasibility of using machine learning techniques, in particular neural networks, for classification of the experimental data from the joint $^\car$C(\textit{n,p}) and $^\car$C(\textit{n,d}) reaction cross section measurement from the neutron time of flight facility n\_TOF at CERN. Each relevant \mbox{$\Delta E$-$E$} pair of strips from two segmented silicon telescopes is treated separately and afforded its own dedicated neural network. An important part of the procedure is a careful preparation of training datasets, based on the raw data from Geant4 simulations. Instead of using these raw data for the training of neural networks, we divide a relevant 3-parameter space into discrete voxels, classify each voxel according to a particle/reaction type and submit these voxels to a training procedure. The classification capabilities of the structurally optimized and trained neural networks are found to be superior to those of the manually selected cuts.

%The implications for the analysis of the experimental data are also discussed.

\end{abstract}

\end{frontmatter}

\section{Introduction}
\label{introduction}

% in 2018

Motivated by an earlier integral cross section measurement of the $^{12}$C(\textit{n,p}) reaction \cite{carbon_prc,carbon_epja}, an energy-differential measurement of the $^\car$C(\textit{n,p}) and $^\car$C(\textit{n,d}) reactions \cite{carbon_prop} was performed at the neutron time of flight facility n\_TOF at CERN. n\_TOF is a sophisticated neutron production facility providing a highly luminous neutron flux spanning 12 orders of magnitude in energy, from 10 meV to 10 GeV. Neutron production, based on the spallation of Pb nuclei from a massive lead spallation target, is induced by 20 GeV proton beam from the CERN Proton Synchrotron. The pulsed beam, with average repetition rate of 0.4~Hz, allows the time of flight technique to be employed for determination of the neutron energy dependence of the measured neutron-induced reactions. Operating since 2001, n\_TOF facility is currently in the third phase of its operation (n\_TOF-Phase3), characterized by parallel utilization of the two experimental areas, referred to as EAR1 and EAR2. Each experimental area is designed in response to a specific set of challenges in measuring the neutron-induced reactions. EAR1 addresses the requirements of the high-energy-resolution as well as the high-neutron-energy measurements, thanks to the long horizontal distance of 185~m from the spallation target, that ensures a long time of flight and an excellent neutron energy resolution. EAR2, with a vertical flight path of 20~m above the same spallation target, is characterized by a significantly increased neutron flux, relative to EAR1, thus being ideal for measurements of reactions with low cross sections and measurements with small and/or highly radioactive samples. A general description of the n\_TOF facility and EAR1 in particular can be found in Ref.~\cite{ntof}. Detailed characteristics of EAR2 are well documented in Refs.~\cite{ear2_1,ear2_2,ear2_3}. We also refer the reader to the in-depth description of the neutron flux evaluation in both experimental areas \cite{flux_ear1,flux_ear2} and the concise overview of experimental activities at n\_TOF \cite{ntof_rev}.

\begin{figure}[t!]
\centering
\includegraphics[width=0.7\linewidth]{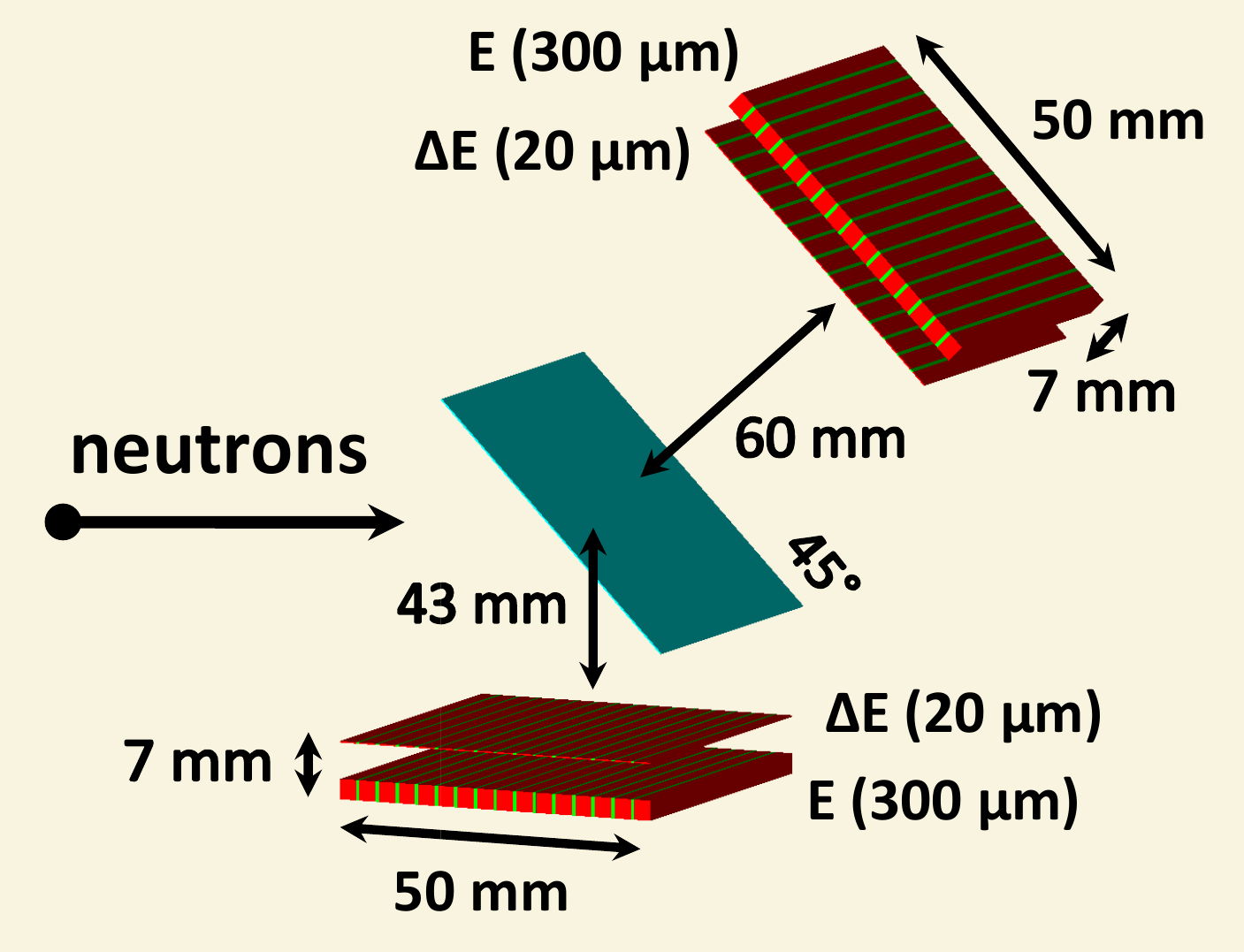}
\includegraphics[width=0.7\linewidth]{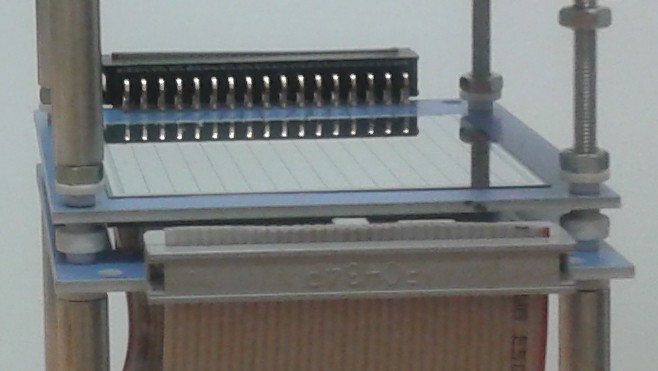}
\caption{Top: schematic diagram of the experimental setup, consisting of two silicon telescopes surrounding the carbon sample (tilted by 45$^\circ$ relative to the neutron beam). Silicon layers are exaggerated in width. Bottom: close-up of a single telescope, showing a stripped structure of a $\Delta E$-layer.}
\label{fig1}
\end{figure}

Experimental setup for the energy-differential measurement of the $^\car$C(\textit{n,p}) and $^\car$C(\textit{n,d}) reactions consists of two silicon telescopes placed outside the neutron beam, surrounding a 0.25~mm thick natural carbon sample. One telescope is parallel to the neutron beam, while the other is parallel to the carbon sample, which itself is tilted by 45$^\circ$ in respect to the beam. This configuration has been specifically optimized in order to maximize a solid angle coverage and minimize systematic effects in data analysis \cite{carbon_prop}. Figure~\ref{fig1} shows a schematic diagram of the experimental setup and a close-up of a single silicon telescope. The excellent charged particle response properties of such telescope configuration are reported in Ref.~\cite{site_np}. In short, each telescope consists of two ($\Delta E$ and $E$) silicon layers -- 20~$\mu$m and 300~$\mu$m thick, respectively, and 7~mm apart. Every layer is \mbox{5 cm $\times$ 5 cm} in lateral dimensions, each comprising 16 silicon strips, 3~mm wide and separated by a thin layer of inactive silicon. The strips in two layers are oriented in the same direction.

%houses

Electronic data from two telescopes (64 channels in total) were digitally recorded at 125~MS/s sampling rate with a 14-bit resolution and analyzed offline by dedicated pulse shape fitting procedures described in Ref.~\cite{psa}. The shape of the electronic signals is reported in Refs.~\cite{site_np,sync}.

One telescope has already been successfully used in an extremely challenging measurement of the $^7$Be(\textit{n,p}) reaction at n\_TOF~\cite{be_np}. Due to the central importance to as yet unresolved Cosmological Lithium Problem, this experiment was also accompanied by a measurement of the $^7$Be(\textit{n,$\alpha$}) reaction~\cite{be_na}, performed by using a similar type of a silicon sandwich detector~\cite{sandw_na}.

Unlike the $^7$Be(\textit{n,p}) measurement performed at EAR2, a joint $^\car$C(\textit{n,p}) and $^\car$C(\textit{n,d}) measurement was performed at EAR1, thus relying on a different data acquisition chain. In the latter case some time offsets were observed between separate data acquisition channels, each corresponding to a particular silicon strip. In order to correct for this in the offline analysis, a simple but effective synchronization method has been developed \cite{sync}. Furthermore, due to the challenging features of the $^\car$C(\textit{n,p}) and $^\car$C(\textit{n,d}) reactions -- in particular the sensitivity to excited states in daughter nuclei and the expected anisotropy in angular distributions of the reaction products -- a special method was developed for the analysis of the experimental data, aiming at the optimal extraction of energy-differential cross sections \cite{angular}. What remains is a classification of the experimental counts according to the reaction type (i.e. the reaction products), in order for the above method to be correctly applied to proper datasets. Such classification, its feasibility as well as its optimal implementation, is the subject of this work. We aim to investigate the feasibility of applying machine learning techniques for this purpose and to put forward all the necessary steps for their implementation.

This work is a part of the ongoing efforts to introduce machine learning techniques into a widespread practice at n\_TOF \cite{neural_ntof1,neural_ntof2}, as they naturally lend themselves to a wide variety of the classification and inference problems. Their amenability to physical sciences has long since been recognized, especially within the nuclear and particle physics \cite{data_book}. As such, they are widely used in the experimental neutron physics \cite{neural_neutron1,neural_neutron2,neural_neutron3}, heavy ion collisions \cite{neural_phenix} and are famously adopted in various Higgs boson related analyses \cite{neural_higgs1,neural_higgs2,neural_higgs3}, to name just a few applications.

A beautiful example (unrelated to n\_TOF) of the particle identification using a $\Delta E$-$E$ technique may be found in Ref.~\cite{ede_example}. The general shape of $\Delta E$-$E$ patterns is well understood from a theoretical point of view~\cite{bethebloch}. Expectedly, one can find earlier applications of neural networks for the particle identification in the silicon-detector based measurements~\cite{site_neural1,site_neural2}. A quite interesting example is a neural network based method that does not require learning~\cite{site_neural3}. Understandably, our analysis features its own specific requirements, warranting a careful documentation of issues and used methods. Of course, neural networks have also been used for the particle identification by means other than $\Delta E$-$E$ discrimination, using the detectors different from silicon-based ones. One recent example may be found in Ref.~\cite{track_neural}.

Section~\ref{motivation} lays out the motivation for adopting machine learning techniques.  Section~\ref{preparation} describes the preparation of the training datasets. Section~\ref{training} reports the details of the neural network optimization and examines the quality of the obtained classification. Section~\ref{conclusions} sums up the main conclusions of this work.
 
%for the classification of data from the joint $^\car$C(\textit{n,p}) and $^\car$C(\textit{n,d}) measurement.

\section{Motivation}
\label{motivation}

In the analysis of the $^\car$C(\textit{n,p}) and $^\car$C(\textit{n,d}) data recorded by two silicon telescopes, each relevant \mbox{$\Delta E$-$E$} pair of silicon strips must be separately taken into account. The relevance of a particular pair of strips is established on a basis of the rate of signals in coincidence (within a time window of $\pm100$~ns \cite{sync}) caused by the detection of charged particles passing through that particular pair. These pairs mostly include the closest and next-to-closest strips from the opposing $\Delta E$ and $E$ layers of a given silicon telescope. In a context of a joint $^\car$C(\textit{n,p}) and $^\car$C(\textit{n,d}) measurement these particles include protons, deuterons, tritons and $\alpha$-particles from different types of neutron-induced reactions on carbon. In that, $\alpha$-particles are of no importance to this work as they are not the subject of current experimental investigation and are well separated from other particle patterns in the coincidence spectra.

Among procedures to be separately applied to each relevant \mbox{$\Delta E$-$E$} pair of strips is a particle recognition, i.e. an identification of a charged particle causing a coincidence of signals in the silicon strips. This is done by observing a distinct spectral signature of specific particles. Based on their mass and charge, each particle type yields separate and recognizable correlations between energies ($\Delta E$ and $E$) deposited in two silicon layers.  The spectral \mbox{$\Delta E$-$E$} patterns also depend on the energy $E_n$ of incident neutrons. A mechanism behind this dependence is clear: the neutron energy $E_n$ determines the energy of released reaction products, thus affecting both the details of their interaction with the silicon detectors and the available energy to be deposited between the silicon layers. Each coincidence is therefore characterized by three parameters -- energies $\Delta E$ and $E$ deposited in two silicon layers, and the neutron energy $E_n$ -- requiring a characterization of the particle-specific patterns in a 3-dimensional parameter space. In that, the neutron energy $E_n$ is inferred from the neutron time of flight, extracted from the signal time relative to reference time provided by a so-called $\gamma$-flash \cite{ntof,sync}.

%Taking advantage of these correlations is the very working principle behind the operation of the silicon telescope.

Although similar between different \mbox{$\Delta E$-$E$} pairs of strips, these patterns are by no means identical. One of the reasons for these discrepancies is geometric in nature: particles incident on different pairs of strips leave the sample under different exit angles (affecting the path-length through the sample) and impinge on pairs of strips at angles dependent on the strip position (affecting the path-length through given strips). The other reason is kinematic: pairs of strips at different positions around the sample intercept the reaction products with different angle-energy correlations due to their kinematic boost from the center of mass frame of the incident neutron and the target nucleus. This affects the correlation between the incident neutron energy $E_n$ and the energies $\Delta E$ and $E$ deposited in different pairs of strips.

Therefore, we are faced not only with the task of fencing the portions of a 3-parameter space ($\Delta E,E,E_n$) corresponding to different particle types -- i.e. of identifying the optimal 2-dimensional boundaries separating these parameter subspaces -- but also of doing that separately for each relevant pair of silicon strips. Adjusting these boundaries manually with hope of achieving an \textit{optimal} separation is a daunting challenge even for a single pair of silicon strips, let alone for a multitude of them. Fortunately, machine learning techniques are an appropriate toll for such task. They excel at complex classifications that would otherwise require a massive manual effort, such as having to identify a proper (analytical or otherwise) form for a multidimensional boundary within a multidimensional parameter space, where the separation of data into desired classes might not even be easily visualized. On the other hand, machine learning techniques usually require a careful preparation and advance investigation of how well they can applied to a task at hand; it is by no means guaranteed that they can yield an optimal or even satisfactory solution to a particular problem. In this work we describe these preparatory considerations and analyze the applicability of neural networks to our particle classification problem. Many of these preliminary procedures -- e.g. preparing a quality training dataset by means of detailed simulations and implementing the methods for assessing the quality of particle classification -- need to be performed regardless of a selected approach to a problem. This will soon become clear, as we will also attempt a simple manual classification in order to judge it against the results obtained from the trained neural networks and to justify a selection of one procedure over the other. We note that this work presents a self-contained \textit{proof of concept} that machine learning techniques may be used for a high-fidelity classification of data obtained by silicon telescopes, with discrimination quality superior to that of the manual classification procedures. The experimental data from a joint measurement of the $^\car$C(\textit{n,p}) and $^\car$C(\textit{n,d}) reactions are pending the completion of several analysis procedures, required before the application of classification techniques described herein.

\section{Training dataset preparation}
\label{preparation}

\subsection{Geant4 simulations}
\label{simulations}

We use Geant4 simulations of the experimental setup, comprising a realistic software replica of the natural carbon sample and of two silicon telescopes, in order to obtain the 3-dimensional coincidence spectra for all relevant $\Delta E$-$E$ pairs of strips, and for each relevant particle type (\textit{p}, \textit{d}, \textit{t}) from the separate neutron induced reactions. As the natural carbon consists of 98.9\% of $^{12}$C and 1.1\% of $^{13}$C, for each of these isotopes we consider four types of reactions -- (\textit{n,p}), (\textit{n,np}), (\textit{n,d}) and (\textit{n,t}) -- making a total of eight reactions. $Q$-values of these reactions and their energy thresholds in the laboratory frame, where the carbon target is at rest, are listed in Table~\ref{tab1}.

Angular distributions of these reactions do not affect the basic shapes of the 3-dimensional coincidence patterns, as these shapes are determined by the kinematics of reaction products and their interaction with the materials in their path. In other words, a spectral place of a given coincidence event, i.e. of a triple ($\Delta E,E,E_n$) corresponding to a given particle/reaction, is almost fully (apart from fluctuations related to the particle interaction with surrounding materials) determined by the initial energy and the emission angle of this particular reaction product. However, the angular distributions do affect the heights of these spectra, i.e. the occupancy of specific portions of a 3-parameter space. Since the specific particle/reaction spectra are expected to overlap, at least in part, these occupancies become relevant in determining which particle/reaction not only occupies, but dominates a given section of a parameter space -- a consideration relevant for the later analysis. Thus, some care must be taken not only in implementing the kinematics of simulated reaction products, but also in implementing the statistical distributions governing their emission. The relevant distributions include (1)~the angular distribution of reaction products, (2)~the distribution of energy thresholds (related to the existence of excited states in daughter nuclei), and (3)~the distribution of energy among reaction products when there are more than two products in the exit channel. Naturally, one should use accurate distributions, provided they were available. In case of the neutron induced reactions on carbon isotopes, many of the distributions are poorly known. For this reason we assume the simplest form of these distributions -- uniform whenever possible -- as the most unbiased course of action in the absence of reliable preexisting data. We describe the role of each relevant distribution below.

\begin{table}[t!]
\caption{Relevant neutron induced reactions on the $^{12}$C and $^{13}$C isotopes, characterized by their $Q$-value and the corresponding energy threshold in the laboratory frame, where the carbon target is at rest prior to reaction.}
\centering
\begin{tabular}{cccc}
\hline\hline
\textbf{Reaction} & $\boldsymbol{Q \; [\mathrm{MeV}]}$ & $\boldsymbol{E_\mathrm{thr} \; [\mathrm{MeV}]}$\\
\hline
$^{12}$C($n,p$)$^{12}$B & $-$12.59 & 13.64 \\
$^{12}$C($n,np$)$^{11}$B & $-$15.96 & 17.30 \\
$^{12}$C($n,d$)$^{11}$B & $-$13.73 & 14.89 \\
$^{12}$C($n,t$)$^{10}$B & $-$18.93 & 20.52 \\
$^{13}$C($n,p$)$^{13}$B & $-$12.65 & 13.64 \\
$^{13}$C($n,np$)$^{12}$B & $-$17.53 & 18.89 \\
$^{13}$C($n,d$)$^{12}$B & $-$15.31 & 16.49 \\
$^{13}$C($n,t$)$^{11}$B & $-$12.42 & 13.38 \\
\hline\hline
\end{tabular}
\label{tab1}
\end{table}

We simulate each reaction from Table~\ref{tab1} separately. For the initial neutron energy $E_n$ we generate the relevant reaction products: protons, deuterons or tritons. The neutron energies are sampled within a relevant energy range from the reaction threshold up to 30~MeV, several MeV above initially estimated upper limit for the reliable analysis of experimental $^\car$C(\textit{n,p}) and $^\car$C(\textit{n,d}) data \cite{carbon_prop}.  The initial kinematic parameters of reaction products are calculated from the manually implemented (relativistic) kinematics, based on emission angle and emission energy sampled from relevant distributions. In that, we also account for their spatial sampling along the width and depth of the carbon sample, in accordance with the neutron beam profile. % from EAR1 at n\_TOF.

% and calculated

\begin{figure*}[t!]
\centering
\includegraphics[width=0.35\linewidth]{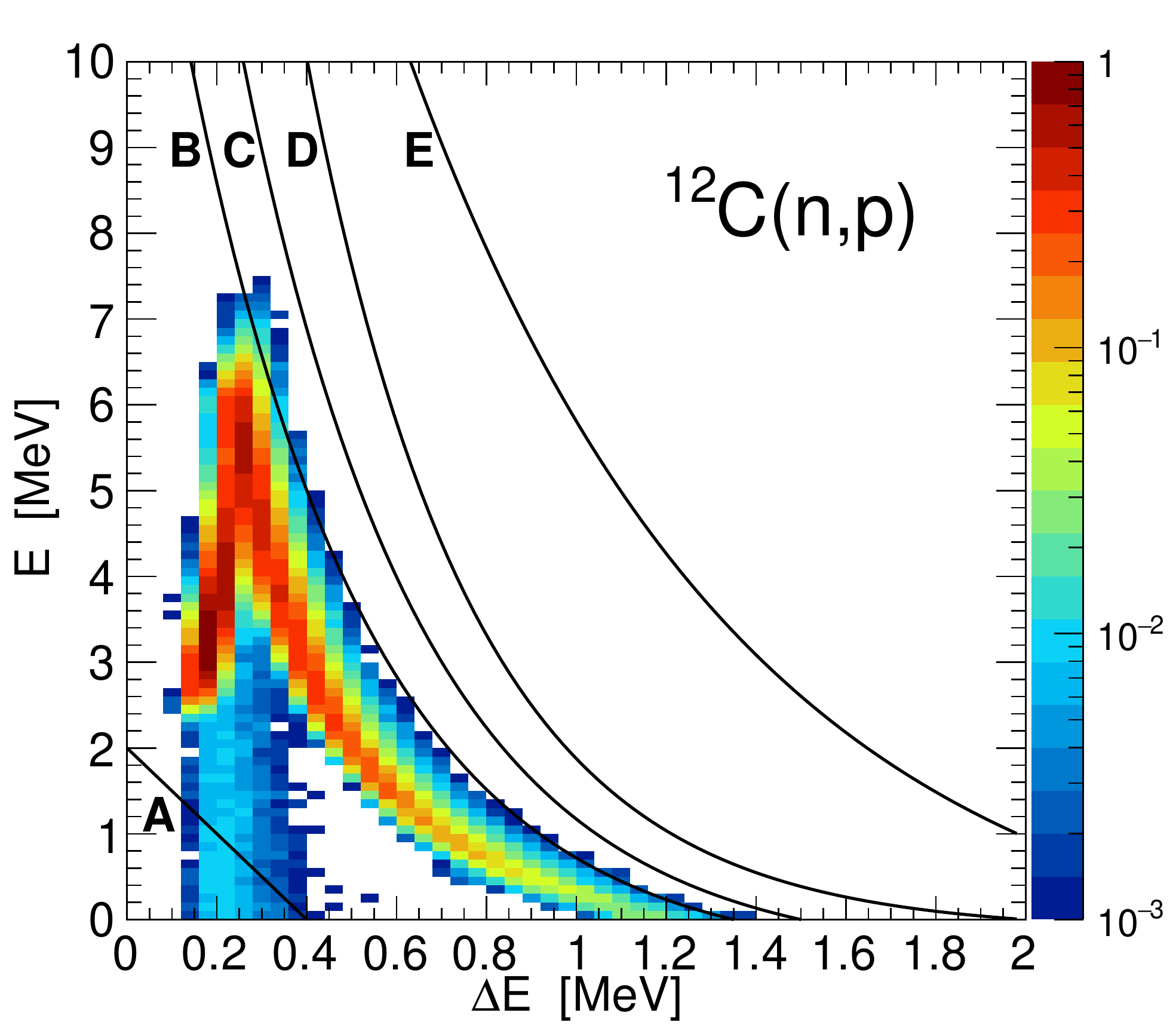}\includegraphics[width=0.35\linewidth]{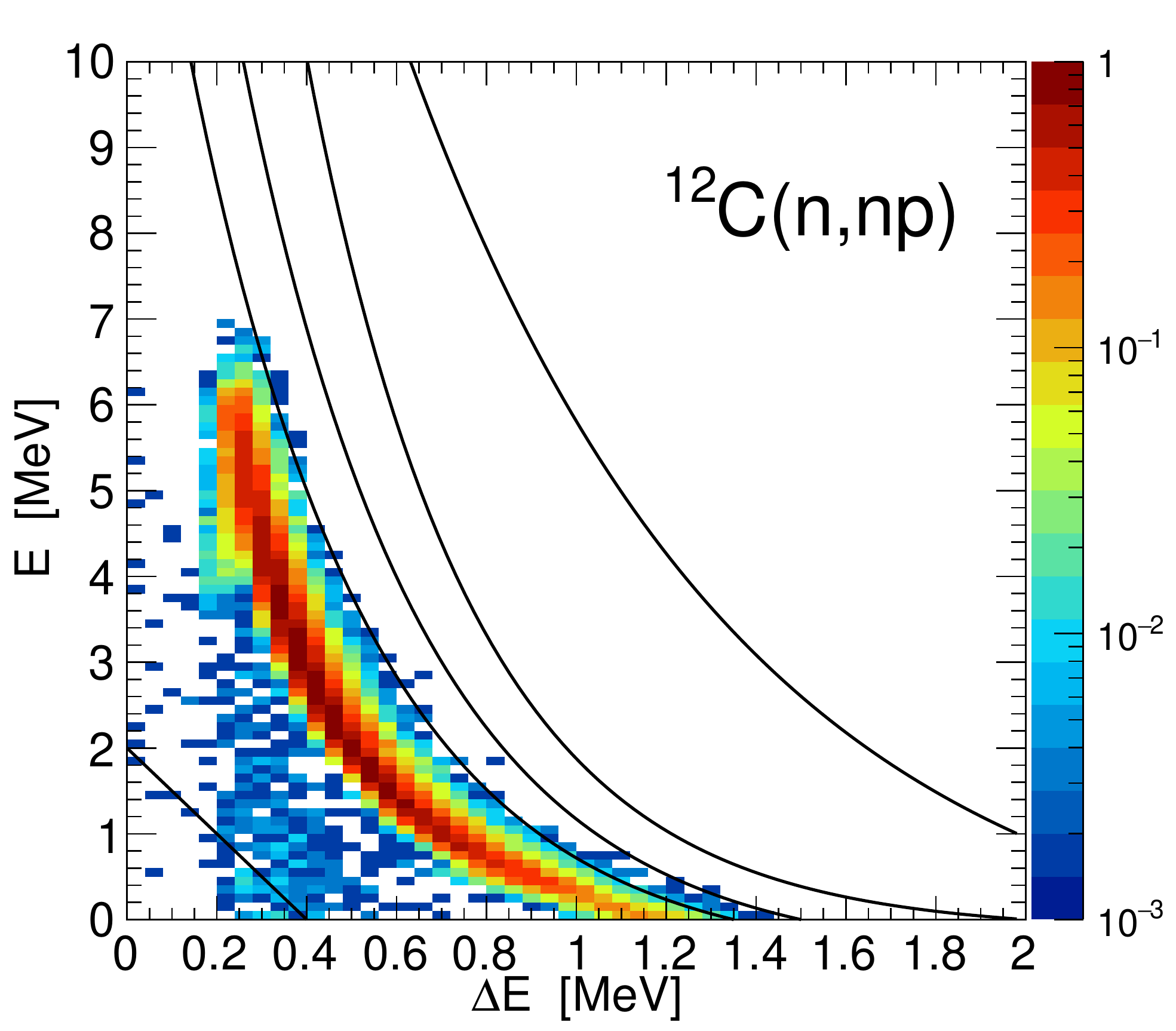}
\includegraphics[width=0.35\linewidth]{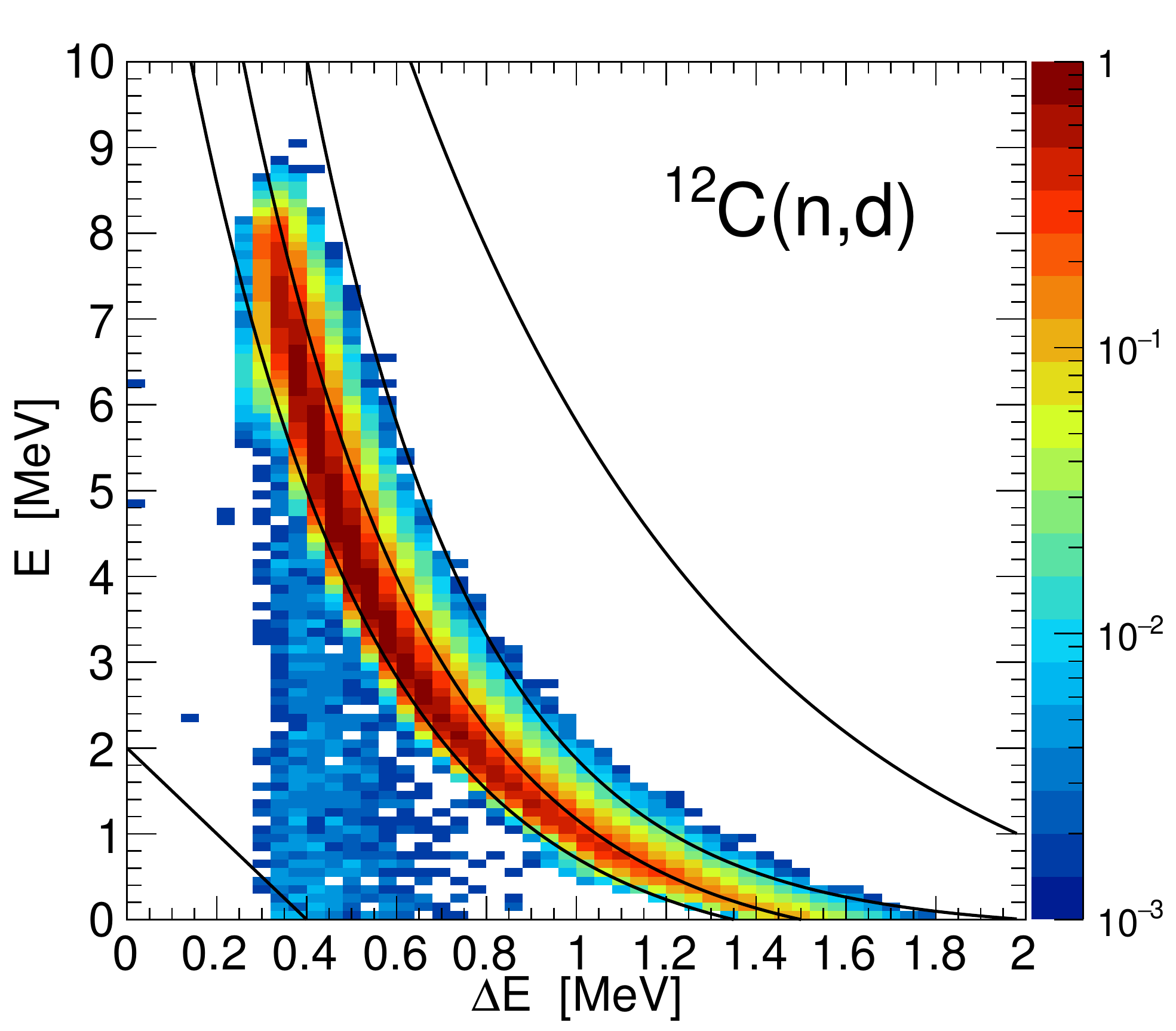}\includegraphics[width=0.35\linewidth]{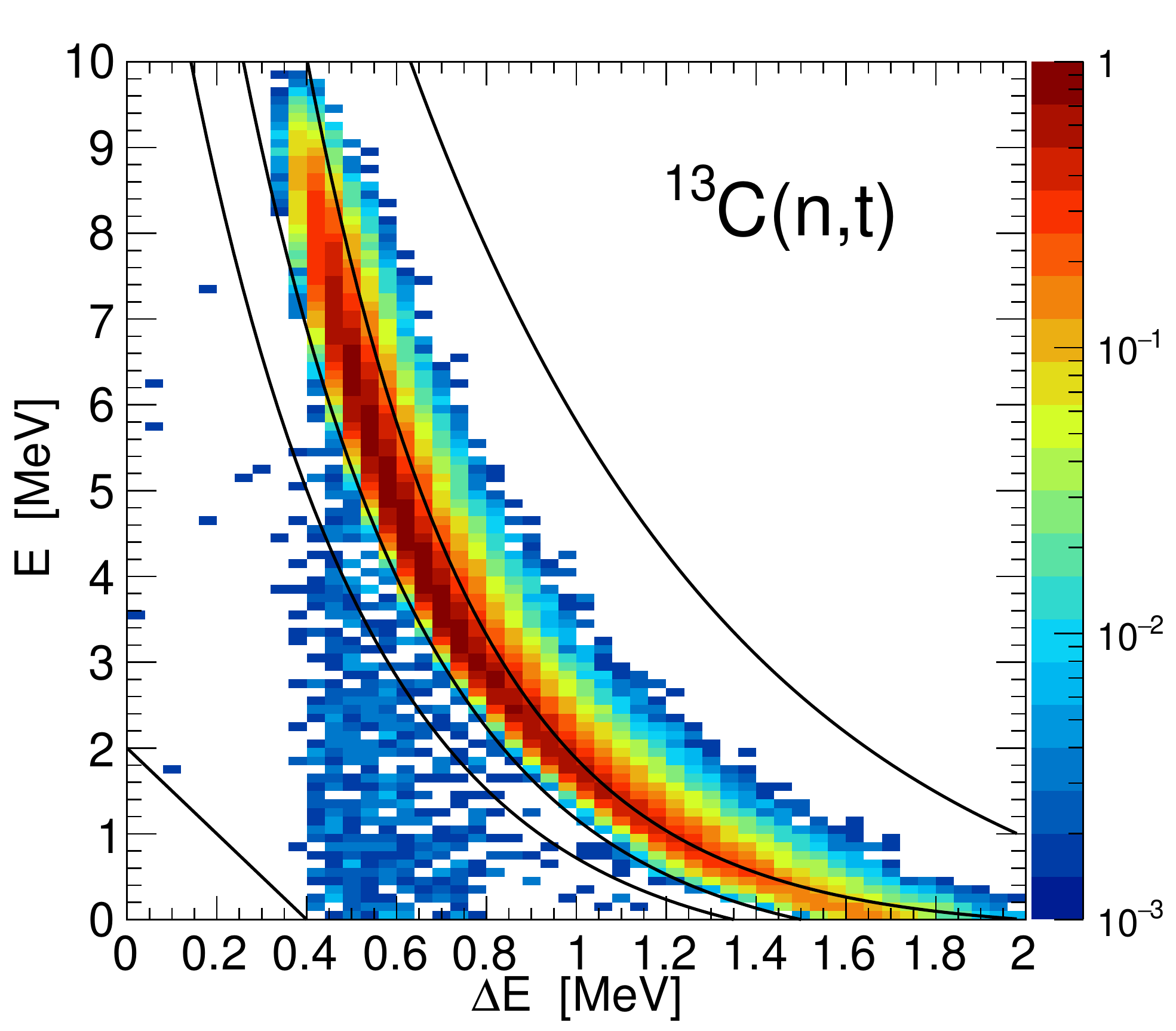}
\caption{Simulated $\Delta E$-$E$ patterns for an arbitrarily selected pair of silicon strips, showing a detector response to protons from the $^{12}$C(\textit{n,p}) and $^{12}$C(\textit{n,np}) reactions, to deuterons from the $^{12}$C(\textit{n,d}) and to tritons from the $^{13}$C(\textit{n,t}) reaction, at a neutron energy of \mbox{$E_n=27$ MeV}. All spectra are smeared by the energy resolution of silicon strips and are normalized such that the highest-content pixel is set to 1. The meaning of displayed manual cuts is discussed in the main text.}
\label{fig2}
\end{figure*}

All daughter nuclei from Table~\ref{tab1} -- $^{10}$B~\cite{b10_states}, $^{11}$B~\cite{b11_states}, $^{12}$B~\cite{b12_states} and $^{13}$B~\cite{b13_states} -- feature a rich spectrum of excited states. As a consequence, each listed reaction may leave a daughter nucleus either in the ground state or in any of its available excited states. For a nucleus left in an excited state, the $Q$-value is lower and the energy threshold is higher than those from Table~\ref{tab1}, which correspond to a daughter nucleus left in the ground state. Furthermore, due to considerable overlap of many excited states because of their large energy widths, there is a continuum of energy thresholds, rather than just a few discrete values. In the absence of detailed information on branching ratios of these states, it would make little sense trying to implement other details of these states in the sampling of reaction products. Instead, after sampling the initial neutron energy $E_n$, we uniformly sample a reaction threshold between the lowest threshold from Table~\ref{tab1} and the available energy $E_n$, corresponding to a selection of a daughter nuclei state between the ground state and the highest one energetically available. In other words, for a given neutron energy we sample a continuous and uniform distribution of branching ratios.

%for a reaction to "choose among"
%accessible

For the (\textit{n,p}), (\textit{n,d}) and (\textit{n,t}) reactions with two particles in the exit channel, the selected neutron energy and the selected reaction threshold uniquely determine the energy of reaction products in the center-of-mass frame. Thus, we have yet to sample only their emission angle. We select an isotropic emission in the center-of-mass frame. By means of manually implemented particle kinematics we boost the kinematic parameters of reaction products into the laboratory frame and generate a relevant particle (\textit{p}, \textit{d}, \textit{t}) with thus obtained energy and emission angle.

For the (\textit{n,np}) reaction with three particles in the exit channel an additional sampling is required. This is due to the fact that for more than two particles in the exit channel, the energy distributions of reaction products are continuous, even in the center-of-mass frame. Again, in the absence of detailed information on these energy distributions, we uniformly sample a proton energy in the center-of-mass frame, between 0 and the maximum kinematically allowed value. As before, proton's kinematic parameters are then boosted into the laboratory frame, and a proton is generated.

%================================

In order to investigate a possibility of background counts from other reactions besides (\textit{n,np}) and (\textit{n,t}), we have run an independent batch of Geant4 simulations, generating incident neutrons and observing a detection of reaction products from three different inelastic cascade models: Binary cascade, Bertini cascade and INCL++/ABLA model (INCL intranuclear cascade coupled to the ABLA deexcitation model). Excluding the counts from the  (\textit{n,np}) and (\textit{n,t}) reactions, the background from other reactions was found completely negligible within a relevant portion of a 3-parameter space, i.e. it is well separated from the patterns of the (\textit{n,p}) and (\textit{n,d}) reactions.

%================================

Figure~\ref{fig2} shows examples of simulated \mbox{$\Delta E$-$E$} patterns for different particle/reaction types, corresponding to the neutron energy of \mbox{$E_n=27$ MeV} -- an arbitrarily selected slice through a 3-parameter space, from an arbitrarily selected \mbox{$\Delta E$-$E$} pair of silicon strips. All simulated data have been smeared by the energy resolution of silicon strips -- relative resolution of 3\% FWHM for $\Delta E$-strips and 0.5\% FWHM for $E$-strips \cite{site_np}. Proton spectra of the (\textit{n,p}) and (\textit{n,np}) reactions and a deuteron spectrum of the (\textit{n,d}) reaction are shown for $^{12}$C. For visual reasons related to a high energy threshold of the $^{12}$C(\textit{n,t}) reaction (see Table~\ref{tab1}), a triton spectrum of the (\textit{n,t}) reaction is shown for $^{13}$C. The patterns of remaining particle/reaction types are equal in shape between $^{12}$C and $^{13}$C. For display purposes all spectra are normalized such that the highest-content pixel is set to~1.

Solid lines from Fig.~\ref{fig2} show the manual cuts that will later be judged against trained neural networks. They also serve as a guide for the eye in a visual analysis of spectra from separate plots. A unique set of manual cuts was determined for all \mbox{$\Delta E$-$E$} pairs of silicon strips, by observing the projected spectra of all counts, from all pairs of silicon strips and from all neutron energies. It is, of course, an entirely fortunate circumstance that such unique set of cuts can be identified at all (at least approximately), implying that the variation of these cuts across the neutron energy and among silicon pairs is not drastic. However, it is expected that this simplified set should not be optimal. It is precisely the goal of machine learning techniques to alleviate both the technical limitations of manual cuts -- such as the necessity for simple analytic forms, preferably independent of $E_n$ -- and to obviate the manual repetition of finding the cuts for each pair of silicon strips separately.

Manual cuts from Fig.~\ref{fig2} are clearly labeled in the first plot, that of the $^{12}$C(\textit{n,p}) reaction. Beyond cuts A and E the amount of data from relevant reactions is negligible or entirely nonexistent. One can mostly find the background counts below cut A. Above cut E is the spectrum of $\alpha$-particles from the (\textit{n,$\alpha$}) reactions, which is well separated from proton, deuteron and triton spectra, hence not being of interest to this work. For this reason all data beyond cuts A and E will be manually discarded, both from neural networks training procedure and from analysis of the experimental data. Cut B separates proton data from deuteron data. Above cut D the amount of deuteron counts is negligible. Cut C was visually estimated as an optimal separation boundary between deuteron and triton spectra. The ordering of cuts C and D indicates a certain overlap between deuteron and triton spectra, clearly demonstrating a necessity of finding an \textit{optimal} separation boundary between them.

%\pagebreak

\subsection{Training approach}

After obtaining the sets of counts corresponding to specific particle/reaction types we are, at least in principle, ready to apply classification methods directly to these raw data, on a count-by-count basis. However, there are several technical issues to consider. The first one is the expected processing time required for training the neural networks, as the amount of counts to be taken for a reliable training should be quite large. At the same time, training needs to be repeated for each relevant pair of silicon strips separately. In that, we will consider 72 pairs of strips in this work, consisting of the relevant closest and next-to-closest $\Delta E$-$E$ neighbors. Though training time issues are not crucial to the quality of training results, it is certainly desirable to reduce this time as much as possible and even to make it independent of a considered amount of counts.

%, selected according to their detection efficiency. In that, we take into account only the pairs with sufficient (arbitrarily selected) number of detected counts. 

The second issue, of much greater impact on the quality of results, is related to a statistical significance of specific counts. The first, most basic measure of this significance is the amount of counts of a specific type. We wish that separation of $\Delta E$-$E$ patterns be \textit{at all} neutron energies $E_n$ \textit{in equal measure affected} by training data from the corresponding neutron energy. However, despite a uniform $E_n$ sampling the number of detected coincidences may differ significantly between separate $E_n$-slices. This may be due to low energy of reaction products at a given neutron energy or due to their angular distribution affecting the emission rate into a solid angle covered by a specific pairs of silicon strips. Since, at the end of training, an optimized boundary between specific particle/reaction counts will be continuous in a 3-parameter space, a separation of $\Delta E$-$E$ patterns from certain $E_n$-slices may be disproportionately affected by a larger number of counts from other slices. Nevertheless, we do not wish to perform a separate training for each $E_n$-slice, as the separation boundaries would discontinuously fluctuate between slices, while a well-behaved boundary should be continuous throughout the entire parameter space.

For these reasons we take different approach to constructing a training dataset, aiming to solve both issues at once. We divide a relevant portion of a 3-dimensional parameter space into discrete volumetric units -- \textit{voxels}. Based on a statistical significance of specific counts we determine to which particle/reaction type each voxel (dominantly) belongs. Then we submit these voxels, instead of single counts, to a neural network training procedure. This approach evidently solves the training time issue, since (for statistically significant amount of counts) the number of filled voxels is considerably smaller than and approximately independent of the number of counts occupying the parameter space. On the other hand, a statistical significance of counts within and between $E_n$-slices is easily affected by manual manipulations during the voxel-type assignment procedure.

\subsection{Training dataset construction}

A relevant portion of a parameter space that we consider spans from 0~MeV to 2~MeV in $\Delta E$, from 0~MeV to 10~MeV in $E$, and from 10.5~MeV to 30.5~MeV in $E_n$. Other details of its partitioning are listed in Table~\ref{tab2}. Since we expect to analyze the experimental data from joint $^\car$C(\textit{n,p}) and $^\car$C(\textit{n,d}) measurement within the neutron energy windows of 1~MeV \cite{carbon_prop}, we adopt for $E_n$ somewhat finer neutron energy resolution of 3 bins per MeV. In total, we divide a parameter space into $3\times10^5$ voxels (the same partitioning as displayed in Fig.~\ref{fig2}). Only a subset of all these voxels will be submitted to a neural network training procedure.

\begin{table}[t!]
\caption{Parameters defining a relevant portion of a 3-parameter space and its partitioning into voxels.}
\centering
\begin{tabular}{c|cccc}
\hline\hline
 & \textbf{Minimum} & \textbf{Maximum} & \textbf{Bins} & \textbf{Bin width}\\
\hline
$\boldsymbol{\Delta E}$& 0 MeV & 2 MeV & 50 & 0.04 MeV\\
$\boldsymbol{E}$& 0 MeV & 10 MeV & 100 & 0.1 MeV\\
$\boldsymbol{E_n}$& 10.5 MeV & 30.5 MeV & 60 & 0.33 MeV\\
\hline\hline
\end{tabular}
\label{tab2}
\end{table}

In order to determine which particle/reaction type (if any at all) dominates specific voxels, we need to manually estimate each type's \textit{detection yield} within a given voxel, since each particle/reaction was simulated separately. For simplicity of notation, we enumerate voxels by a triple \mbox{$(\Delta E,E,E_n)$} of energies relevant to that voxel (e.g. energy coordinates of its center). In that, we are not interested in absolute detection yields but only in relative contributions between different particle/reaction types. Thus, for each particular type of reaction $\R$ -- (\textit{n,p}), (\textit{n,np}), (\textit{n,d}) or (\textit{n,t}) -- on each particular carbon isotope $\Z$ (either $^{12}$C or $^{13}$C; \mbox{$\Z=12,13$}), detected in coincidence by each particular $\Delta E$-$E$ pair of silicon strips $\pair$, we define a simpler quantity $x_{\Z,\R}^{(\pair)}$ that is proportional to a true detection yield\footnote{
For thin samples, a \textit{reaction yield} (the amount of reaction products per incident neutron) equals \mbox{$Y_{\Z,\R}(E_n)=\eta_\Z\sigma_{\Z,\R}(E_n)$}, with $\sigma_{\Z,\R}(E_n)$ as the reaction cross section and \mbox{$\eta_\Z=A_\Z\eta$} as an areal density of a particular carbon isotope (in number of atoms per unit area), $A_\Z$ and $\eta$ being an isotopic abundance and a total areal density of carbon sample. A \textit{true detection yield} $X_{\Z,\R}^{(\pair)}$ is then:
\begin{linenomath}\begin{equation*}
X_{\Z,\R}^{(\pair)}(\Delta E,E,E_n)=\epsilon_{\Z,\R}^{(\pair)}(\Delta E,E,E_n)Y_{\Z,\R}(E_n)=\eta\frac{N_{\Z,\R}^{(\pair)}(\Delta E,E,E_n)}{\mathrm{N}_{\Z,\R}(E_n)} A_\Z\sigma_{\Z,\R}(E_n),
\end{equation*}\end{linenomath}
with $\epsilon_{\Z,\R}^{(\pair)}$ as a detection efficiency, corresponding to a fractional term from Eq.~(\ref{det_yield}). A simplified definition of detection yield from Eq.~(\ref{det_yield}) differs only by a factor $\eta$ from true detection yield.
}, up to the areal density of a sample as an omitted multiplicative factor. For simplicity and clarity we still refer to this quantity as \textit{detection yield} throughout the paper. For each voxel we estimate these specific detection yields as:
\begin{linenomath}\begin{equation}
x_{\Z,\R}^{(\pair)}(\Delta E,E,E_n)\approx\frac{N_{\Z,\R}^{(\pair)}(\Delta E,E,E_n)}{\mathrm{N}_{\Z,\R}(E_n)}A_\Z \bar{\sigma}_{\Z,\R}.
\label{det_yield}
\end{equation}\end{linenomath}
\mbox{$\mathrm{N}_{\Z,\R}(E_n)$} is a simulated number of reaction products, generated for the reaction $\R$ on the carbon isotope $\Z$ within the neutron energy window at $E_n$. \mbox{$N_{\Z,\R}^{(\pair)}(\Delta E,E,E_n)$} is a number of reaction products detected within a given voxel of the parameter space by the silicon pair $\pair$. $A_\Z$ are natural abundances of relevant carbon isotopes (\mbox{$A_{12}=98.9\%$} and \mbox{$A_{13}=1.1\%$}). $\bar{\sigma}_{\Z,\R}$ the roughly estimated cross section averages within neutron energy intervals spanning from a particular reaction threshold up to 30~MeV. Since evaluated data on these cross sections show large variations between available libraries -- both in range and in values -- the cross section estimates were obtained from Geant4 simulations, by averaging the cross sections predicted by three different inelastic cascade models: Binary cascade, Bertini cascade and INCL++/ABLA model (INCL intranuclear cascade coupled to the ABLA deexcitation model). Energy dependence of each reaction cross section was first extracted from Geant4 simulations using a method from Ref.~\cite{carbon_epja}, described in \ref{cross_section}. For each reaction an energy dependence was averaged between three cascade models and the averages $\bar{\sigma}_{\Z,\R}$ within the energy intervals up to 30~MeV were estimated. We obtained the following values: \mbox{$\bar{\sigma}_{12,(n,p)}=22$~mb}, \mbox{$\bar{\sigma}_{12,(n,np)}=15$~mb}, \mbox{$\bar{\sigma}_{12,(n,d)}=27$~mb}, \mbox{$\bar{\sigma}_{12,(n,t)}=8.5$~mb}, \mbox{$\bar{\sigma}_{13,(n,p)}=3.3$~mb}, \mbox{$\bar{\sigma}_{13,(n,np)}=15$~mb}, \mbox{$\bar{\sigma}_{13,(n,d)}=16$~mb}, \mbox{$\bar{\sigma}_{13,(n,t)}=14$~mb}. Absolute values of these cross sections are of no importance to this work, only their relative magnitude. In addition, we have confirmed that these values do not play a crucial role in the upcoming analysis of our data, i.e. a voxel classification is rather insensitive to variations in thus obtained $\bar{\sigma}_{\Z,\R}$.

\begin{figure*}[t!]
\centering
\includegraphics[width=0.35\linewidth]{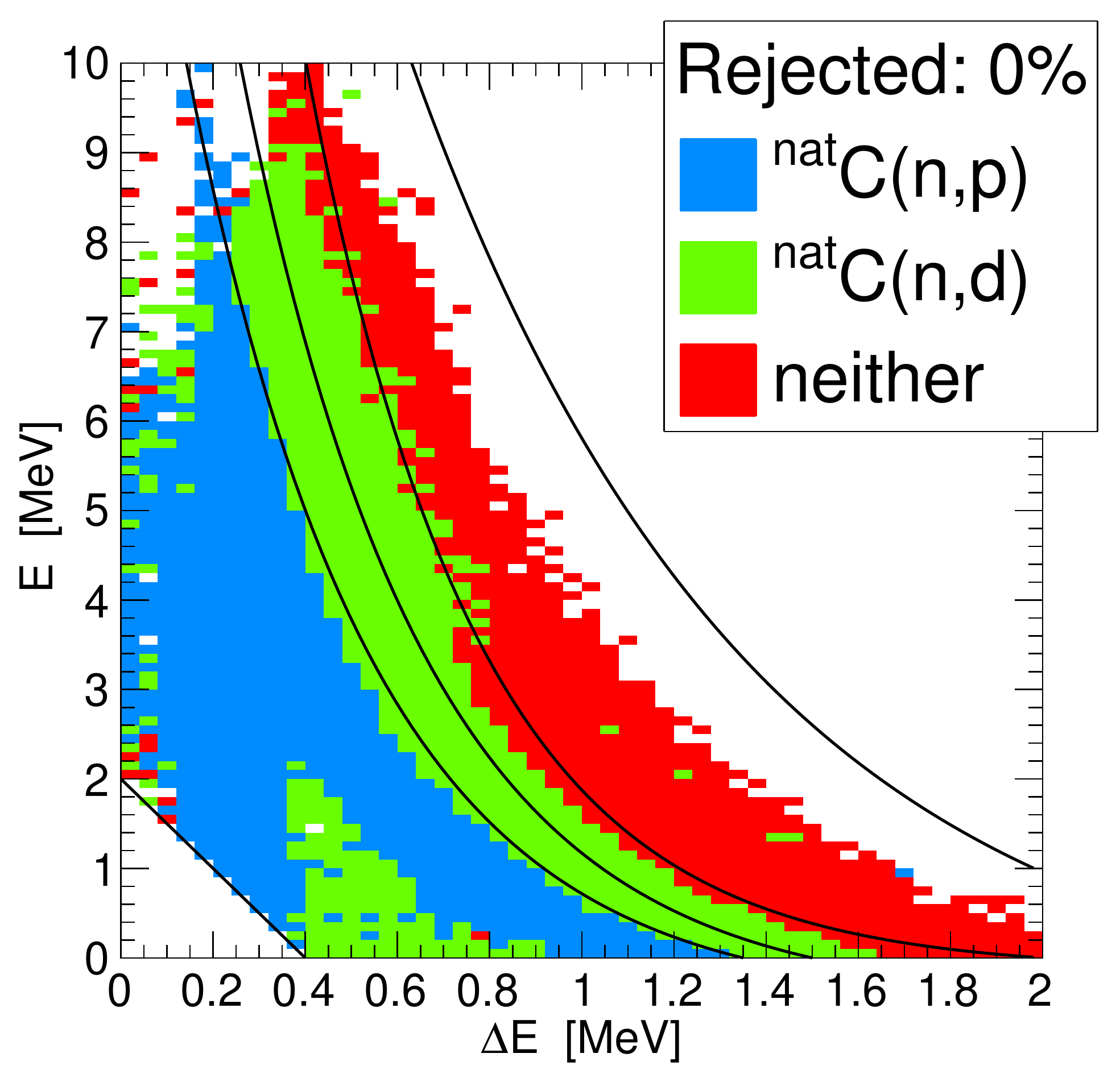}\includegraphics[width=0.35\linewidth]{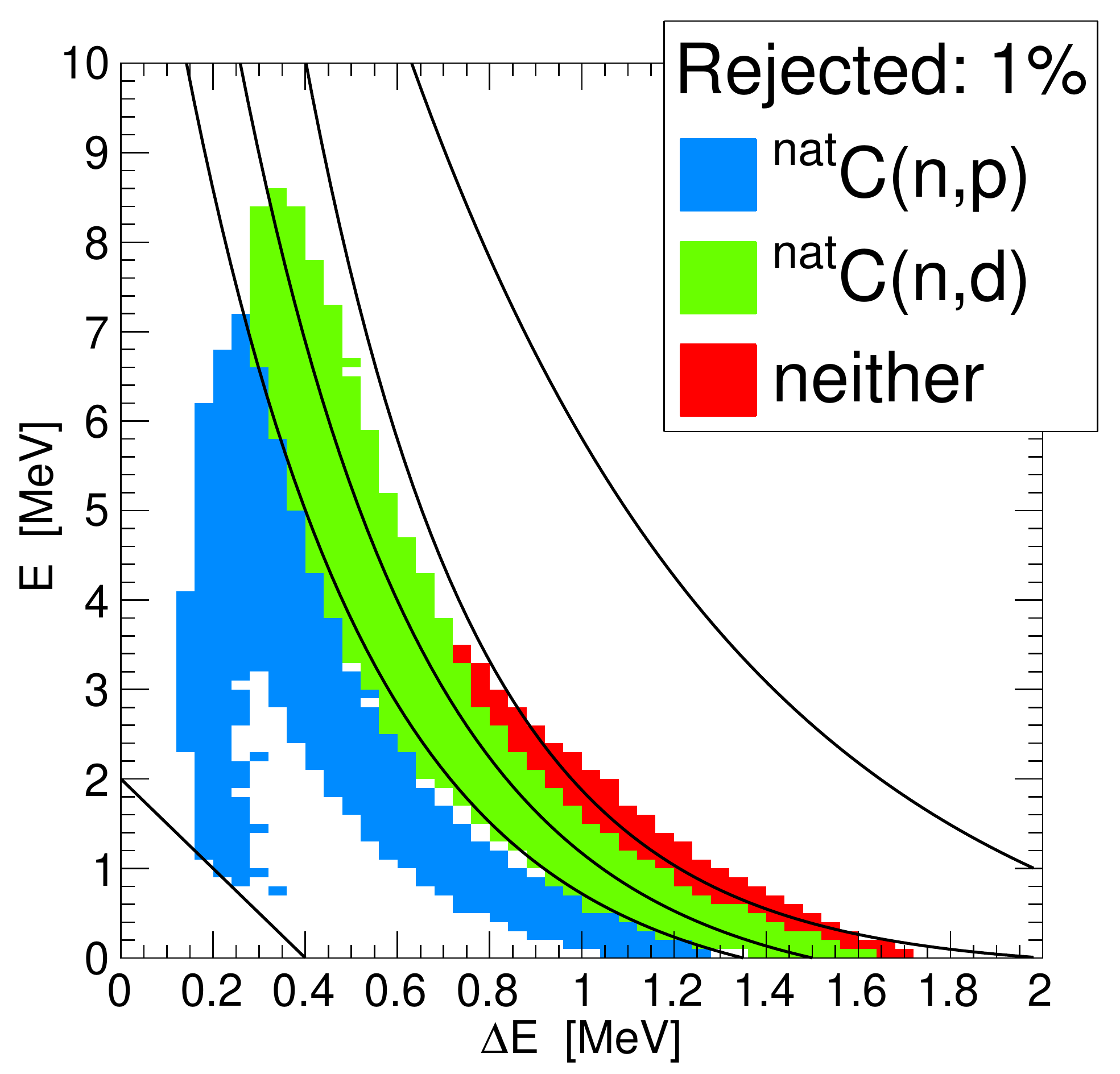}
\caption{Assignment of the reaction types at \mbox{$E_n=27$ MeV}, for an arbitrarily selected pair of silicon strips from Fig.~\ref{fig2}. The types are determined based on the reactions shown in Fig.~\ref{fig2} and four more reactions not shown there: $^{13}$C(\textit{n,p}), $^{13}$C(\textit{n,np}), $^{13}$C(\textit{n,d}) and $^{12}$C(\textit{n,t}). Left panel: all voxels originally determined as of (\textit{n,p}), (\textit{n,d}) or background type (none manually rejected). Right panel: voxels remaining after rejecting 1\% of total detection yield from this particular $E_n$-slice. White voxels are of the undecided type (empty, of unclear classification or rejected).}
\label{fig3}
\end{figure*}

Due to similar reaction thresholds for $^{12}$C and $^{13}$C (Table~\ref{tab1}), reaction products of the same type from two carbon isotopes can not be experimentally distinguished, as they occupy the same portions of a parameter space. Therefore, for each reaction type $\R$ we continue observing a joint contribution from both carbon isotopes, in accordance with their natural abundance. From this point on, each quantity characterized by $\R$ and not by $\Z$ refers to the natural carbon, as used in a joint measurement of the $^\car$C(\textit{n,p}) and $^\car$C(\textit{n,d}) reactions. Joint  contributions to a detection yield follow simply as:
\begin{linenomath}\begin{equation}
x_\R^{(\pair)}(\Delta E,E,E_n)=x_{12,\R}^{(\pair)}(\Delta E,E,E_n)+x_{13,\R}^{(\pair)}(\Delta E,E,E_n).
\end{equation}\end{linenomath}

For each voxel from every silicon pair we now observe the relative contributions $\rel_{\R}^{(\pair)}$ from specific reaction types:
\begin{linenomath}\begin{equation}
\rel_{\R}^{(\pair)}(\Delta E,E,E_n)=\frac{x_{\R}^{(\pair)}(\Delta E,E,E_n)}{\sum_{\R'}x_{\R'}^{(\pair)}(\Delta E,E,E_n)},
\end{equation}\end{linenomath}
and classify the voxels, i.e. assign to each of them one of four labels using the following rules:
\begin{itemize}%\itemsep0em
\item if the relative yield of the (\textit{n,np}) and (\textit{n,t}) reactions exceeds 50\%: \mbox{$\rel_{(n,np)}^{(\pair)}(\Delta E,E,E_n)+\rel_{(n,t)}^{(\pair)}(\Delta E,E,E_n)> 50\%$}, label a voxel as the \textit{background type};
\item if the relative yield of the (\textit{n,p}) reactions exceeds 50\%: \mbox{$\rel_{(n,p)}^{(\pair)}(\Delta E,E,E_n)> 50\%$}, label a voxel as the \textit{(\textit{n,p})-type};
\item if the relative yield of the (\textit{n,d}) reactions exceeds 50\%: \mbox{$\rel_{(n,d)}^{(\pair)}(\Delta E,E,E_n)> 50\%$}, label a voxel as the \textit{(\textit{n,d})-type};
\item otherwise, if empty of counts or not satisfying any of the previous conditions (i.e. if being of unclear classification), label a voxel as the \textit{undecided type}.
\end{itemize}

Left panel from Fig.~\ref{fig3} shows an example of thus assigned types within the $E_n$-slice at 27~MeV, corresponding to the data from Fig.~\ref{fig2} (keep in mind that four other reactions have been taken into account, besides those shown in Fig.~\ref{fig2}). Let us consider the effect on the quality of neural network training, caused by such arrangement of voxel types. There are some sporadic voxels far away from the main pattern corresponding to a specific voxel type, most prominently on a left side of the plot, where several isolated (\textit{n,d}) and background voxels mix with the main (\textit{n,p}) pattern. Furthermore, there is an entire island of (\textit{n,d}) voxels closed off by the (\textit{n,p}) pattern. From Fig.~\ref{fig2} we make two important observations. The first is that these voxels are statistically insignificant, relative to the main pattern of their type. The second is that the \textit{primary} reason for these voxels not to have been assigned the (\textit{n,p}) type is not so much the fact that \mbox{non-(\textit{n,p})} counts from these voxels dominate over the (\textit{n,p}) counts, but rather that there are no (\textit{n,p}) counts in these voxels at all. If the neural networks were trained on such set of voxels, a 3-parameter boundary between the reaction types would be disproportionately affected by these voxels, considering their statistical (in)significance. For this reason we employ a simple method to exclude such insignificant voxels from interfering with training procedure, in order to ensure greater robustness of boundaries between the reaction types. A method consists in rejecting from each $E_n$-slice the voxels with the lowest detection yield, until a predefined portion (specifically, 1\%) of total yield from that particular $E_n$-slice has been removed. Formally stated, for each voxel let \mbox{$x^{(\pair)}(\Delta E,E,E_n)$} be a \textit{joint detection yield} from all reactions:
\begin{linenomath}\begin{equation}
x^{(\pair)}(\Delta E,E,E_n)=\textstyle \sum_{\R}x_\R^{(\pair)}(\Delta E,E,E_n).
\end{equation}\end{linenomath}
Furthermore, for each $E_n$-slice let $\hat{x}_i^{(\pair)}(E_n)$ be an \textit{ordered}, increasing array of joint yields from within the same $E_n$-slice, where index $i$ substitutes $(\Delta E,E)$ pairs of parameters, ordered according to increasing voxel content. The number $\kappa^{(\pair)}(E_n)$ of ordered voxels from a given $E_n$-slice, that we label as  empty, i.e. as the undecided type, is the maximum $\kappa^{(\pair)}(E_n)$ for which it holds:
\begin{linenomath}\begin{equation}
\frac{\sum_{i=1}^{\kappa^{(\pair)}(E_n)} \hat{x}_i^{(\pair)}(E_n)}{\sum_{\Delta E,E} x^{(\pair)}(\Delta E,E,E_n)}\le1\%.
\label{reject}
\end{equation}\end{linenomath}
In other words, we reject at most 1\% of total detection yield from each $E_n$-slice.

%non-empty, NE non-undecided!!!

%in increasing order

%Furthermore, for each $E_n$-slice let $\chi_i^{(\pair)}(E_n)$ be an \textit{ordered}, increasing array of these joint detection yields from non-empty voxels within the same $E_n$-slice. From among the total of $K^{(\pair)}(E_n)$ ordered voxels in a given $E_n$-slice we mark the lowest-content voxels as empty, i.e. as the undecided type. In that, the number of them that we eliminate is maximum $k^{(\pair)}(E_n)$ for which it holds:

Right panel from Fig.~\ref{fig3} shows the result of such elimination of superfluous voxels from the left panel. Statistical insignificance of rejected voxels is to be appreciated, as they make up almost half of all initial voxels, while carrying only 1\% of a statistical norm. %, i.e. of the detection yield.

\section{Neural network training}
\label{training}

We use the neural network training capabilities of \texttt{TMultiLayerPerceptron} class from ROOT. The sets of voxels that are submitted to a training procedure are represented by right panel from Fig.~\ref{fig3}. Input data consist of the triples ($\Delta E,E,E_n$), i.e. of parametric coordinates of each voxel. Output data consist of each voxel's type: (\textit{n,p})-type, (\textit{n,d})-type or background type. Voxels of the undecided type are not submitted to a training procedure at all. Considering the undecided voxels as separate relevant type would overly restrict the boundary between the (\textit{n,p})/(\textit{n,d}) patterns and the empty portions of parameter space, effectively causing the undecided type to behave as the background type. Background type indicates that certain portions of parameter space are decidedly not dominated either by the (\textit{n,p}) or (\textit{n,d}) reaction. On the other hand, the purpose of the undecided type is to allow the generalization and extension of specific type-patterns into the portions of parameter space that are not covered by training data.

Though the default Broyden-Fletcher-Goldfarb-Shanno (BFGS) training method from \texttt{TMultiLayerPerceptron} class performs reasonably well, we have found that the stochastic minimization with manually optimized internal hyperparameters \mbox{$\tau=5$} and \mbox{$\eta=0.01$}, and default values of \mbox{$\eta_\text{decay}=1$}, \mbox{$\delta=0$} and \mbox{$\epsilon=0$} converges faster (in fewer training iterations) and requires only about half of processing time per iteration, in comparison with the BFGS method with default settings. Therefore, in this work we use a stochastic minimization with the listed hyperparameter values. For hidden neurons a sigmoid function was used as an activation function.

\begin{figure}[t!]
\centering
\includegraphics[width=0.7\linewidth]{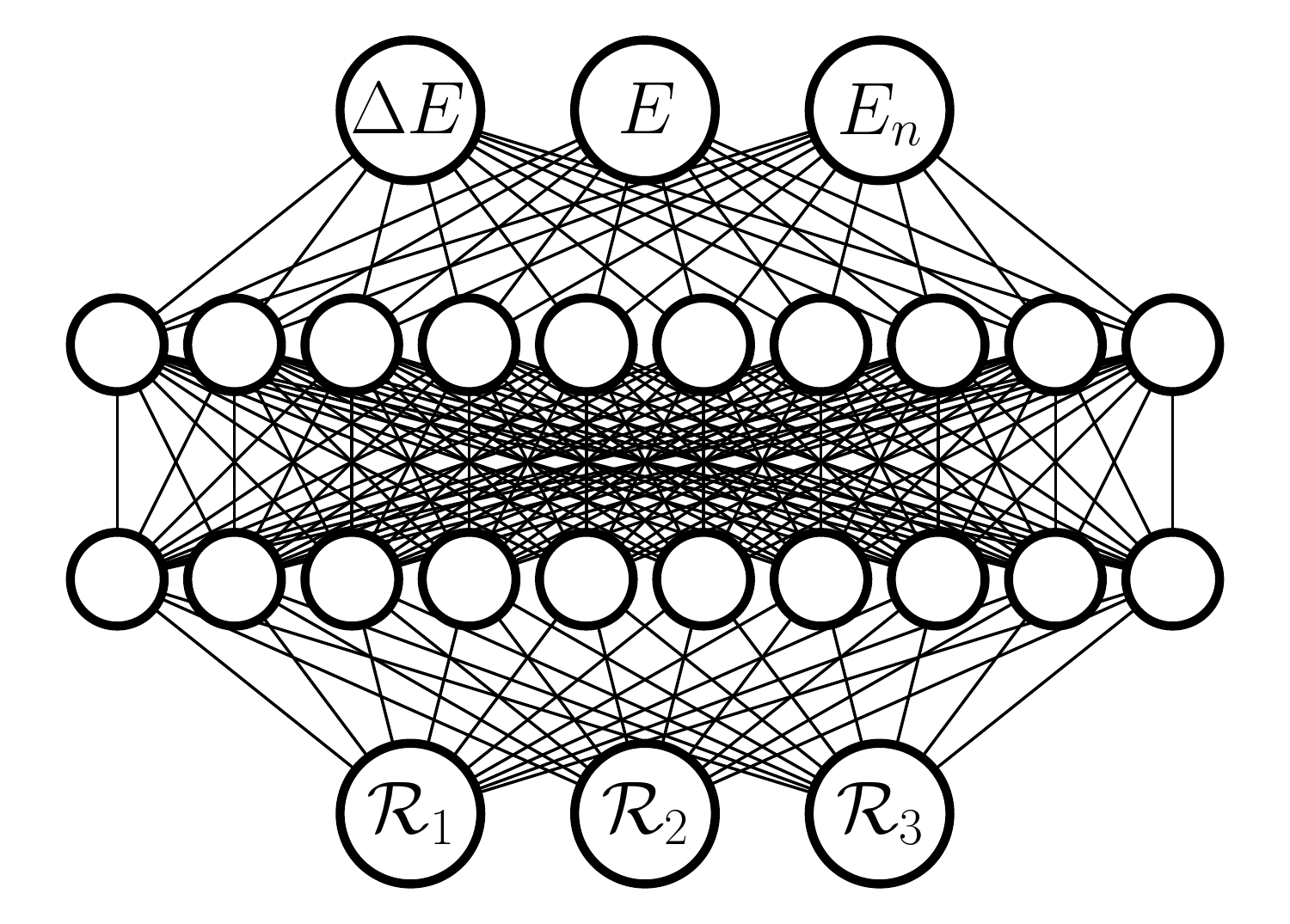}
\caption{Structure of the adopted neural networks for each pair of $\Delta E$-$E$ strips. Inputs are the coordinates \mbox{$(\Delta E,E,E_n)$} of each voxel remaining after the rejection procedure from Eq.~(\ref{reject}). Outputs are reaction labels $\R_1,\R_2,\R_3\in\{0,1\}$. $\R_1$~stands for the $(n,p)$ reaction, $\R_2$ for the $(n,d)$ reaction and $\R_3$ for the background reactions. Since the undecided voxels are excluded from training procedure, each submitted voxel has exactly one of the labels equal to 1 and the remaining ones equal to 0.}
\label{fig4}
\end{figure}

We have found that the optimal network structure consists of 2 layers of neurons, each composed of 10 neurons. An important requirement in identifying this optimized structure was that both the quality and the consistency of trained boundaries remain stable even when introducing variations in the input data, e.g. by varying a rejection level from Eq.~(\ref{reject}). In conclusion, we employ a 2-layer neural network with 10 neurons in each layer. The adopted network structure is displayed in Fig.~\ref{fig4}.

Though of secondary importance, a voxel rejection procedure from Eq.~(\ref{reject}) further reduces the amount of voxels to be submitted to a training procedure (thus further reducing a training time) since the undecided voxels are not submitted at all. For 72 pairs of silicon strips that we consider in this work, consisting of the relevant closest and next-to-closest $\Delta E$-$E$ neighbors, a number of specific-type voxels per silicon pair varies between approximately \mbox{47\,000} and \mbox{87\,600} in case of no voxel rejection (represented by left panel from Fig.~\ref{fig3}). After rejecting 1\% of the lowest voxel content (right panel from Fig.~\ref{fig3}), between \mbox{29\,000} and \mbox{47\,500} voxels remain per pair. These numbers should also be compared against the initial $3\times10^5$ voxels composing an entire parameter space.

\begin{figure*}[t!]
\centering
\includegraphics[width=0.39\linewidth]{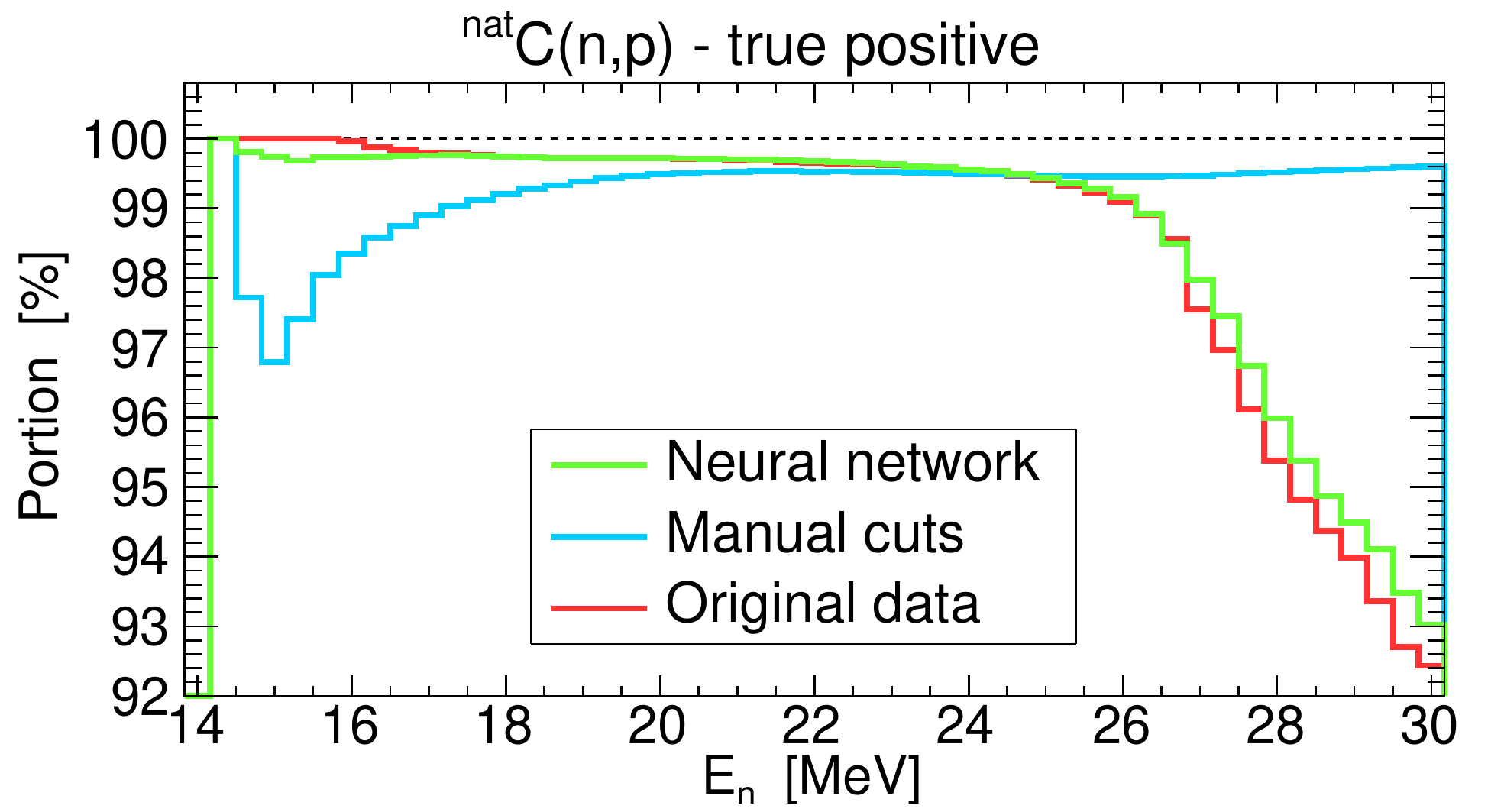}\includegraphics[width=0.39\linewidth]{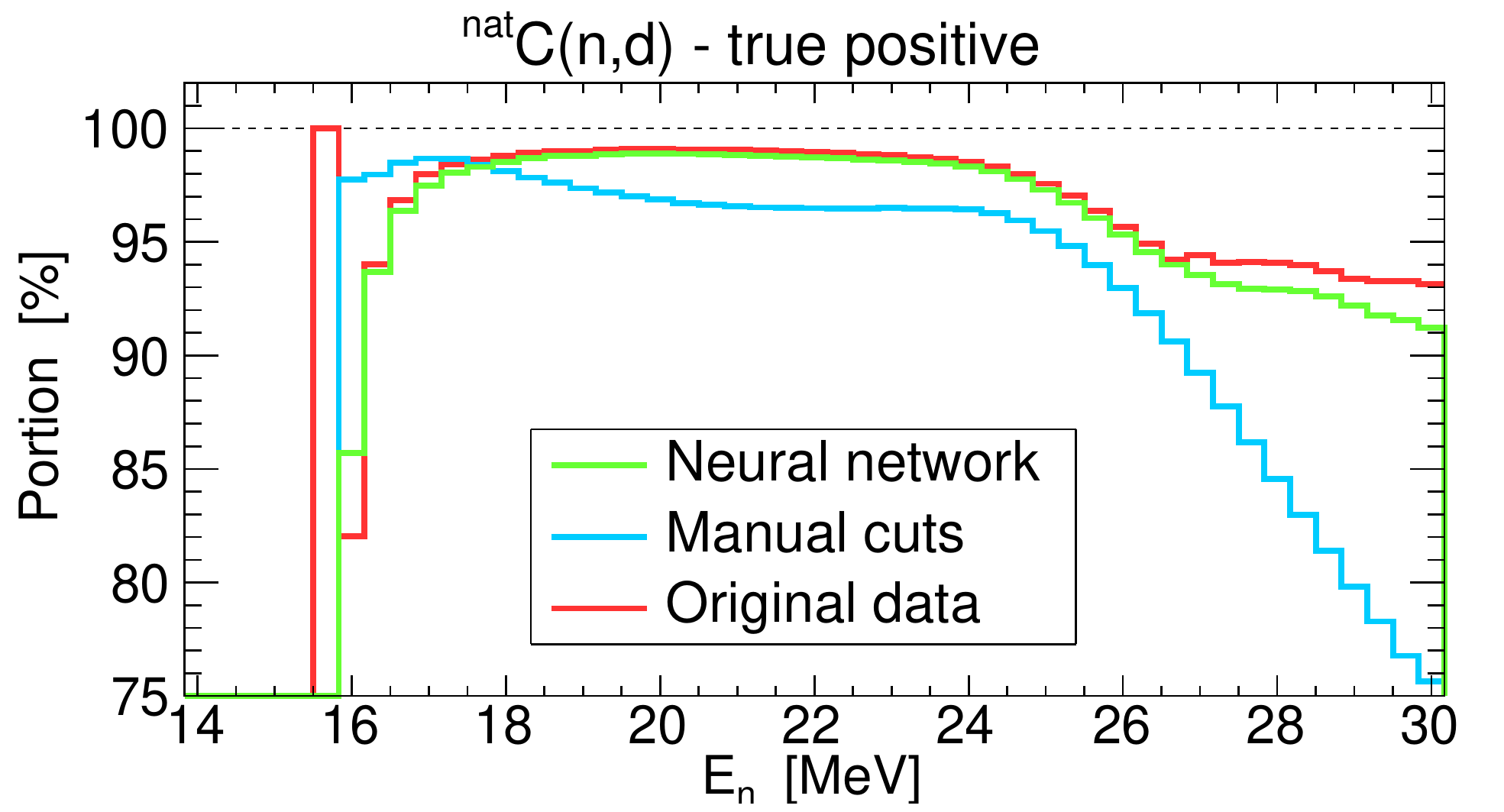}
\includegraphics[width=0.39\linewidth]{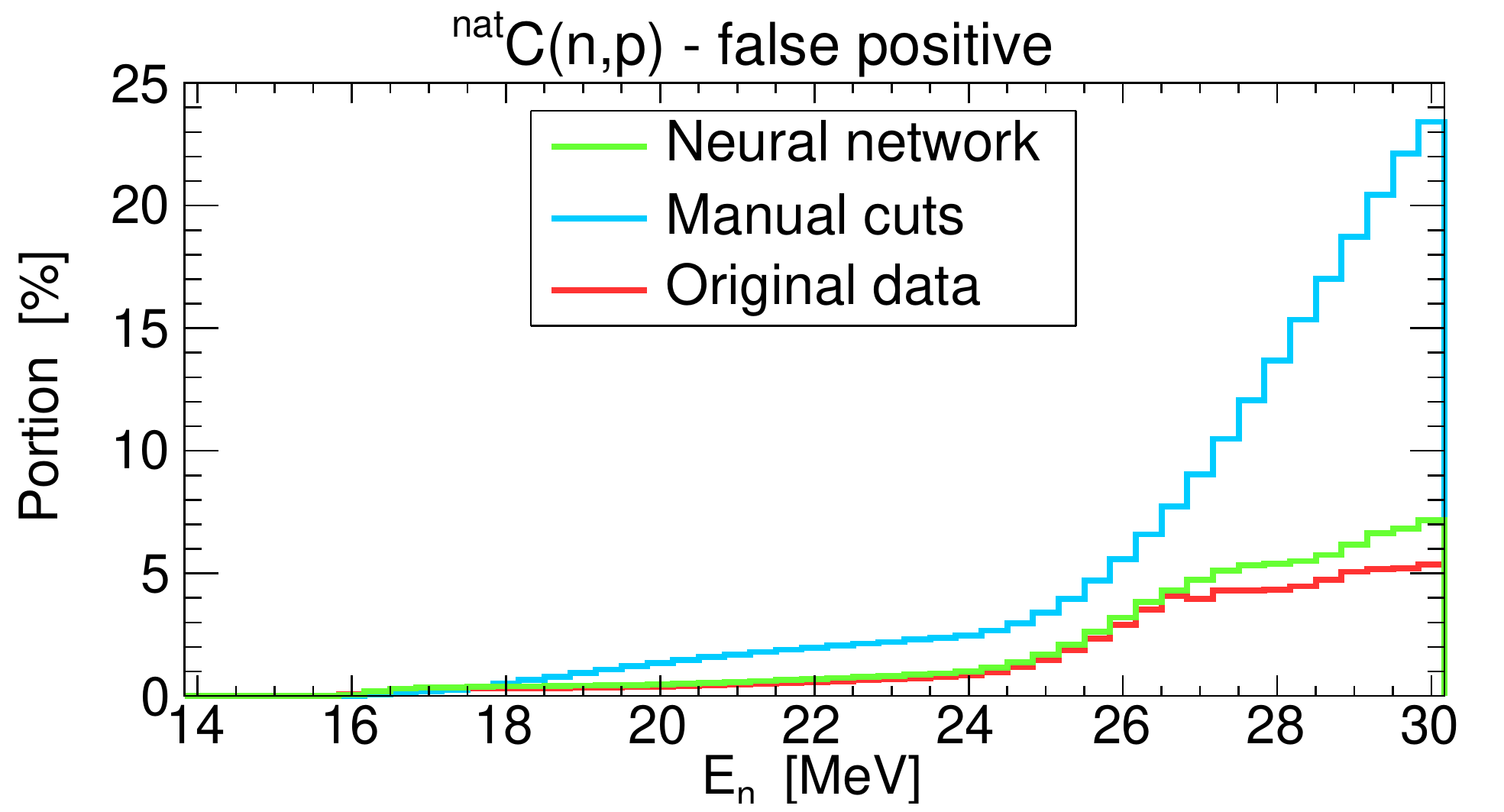}\includegraphics[width=0.39\linewidth]{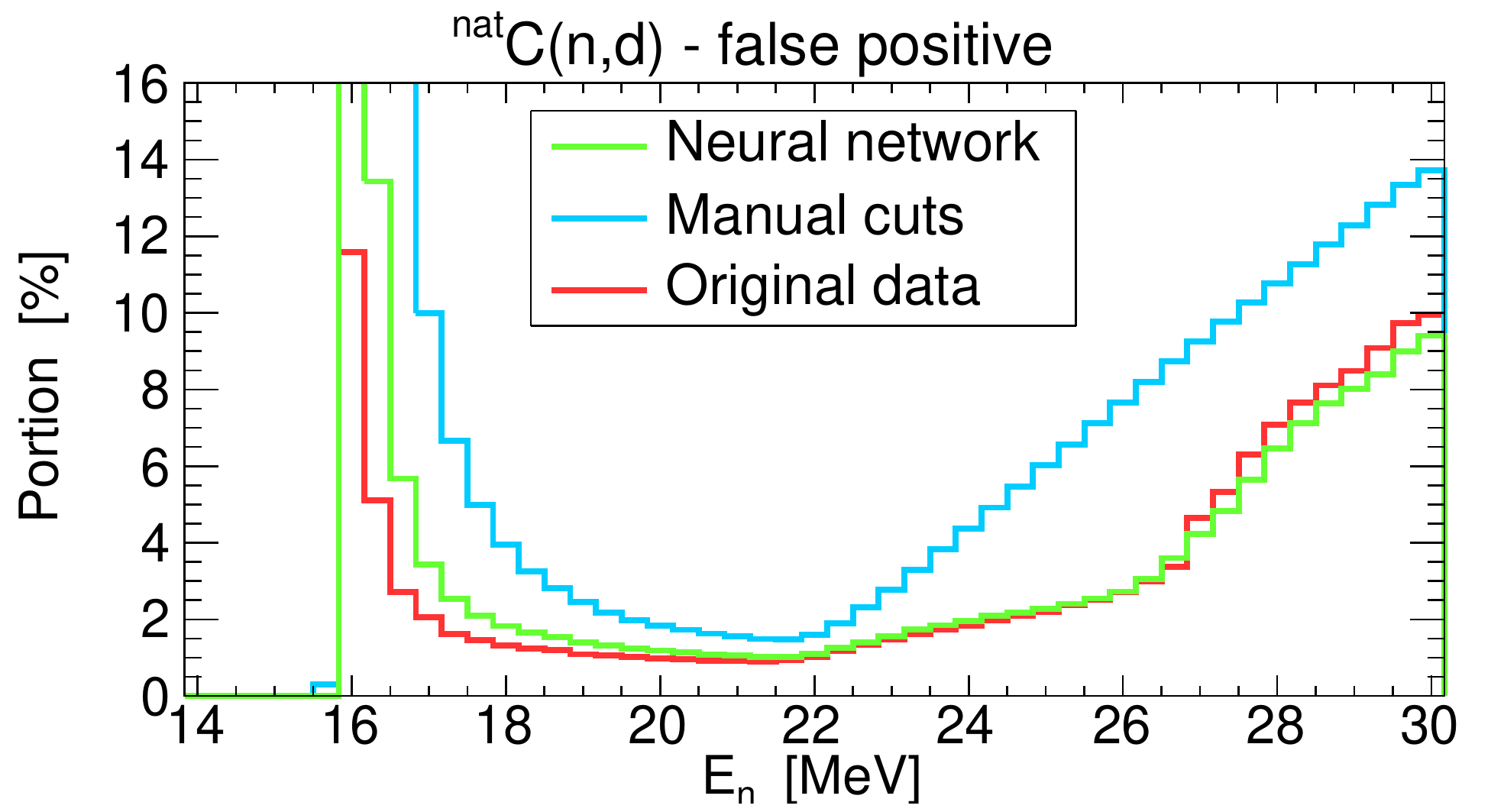}
\includegraphics[width=0.39\linewidth]{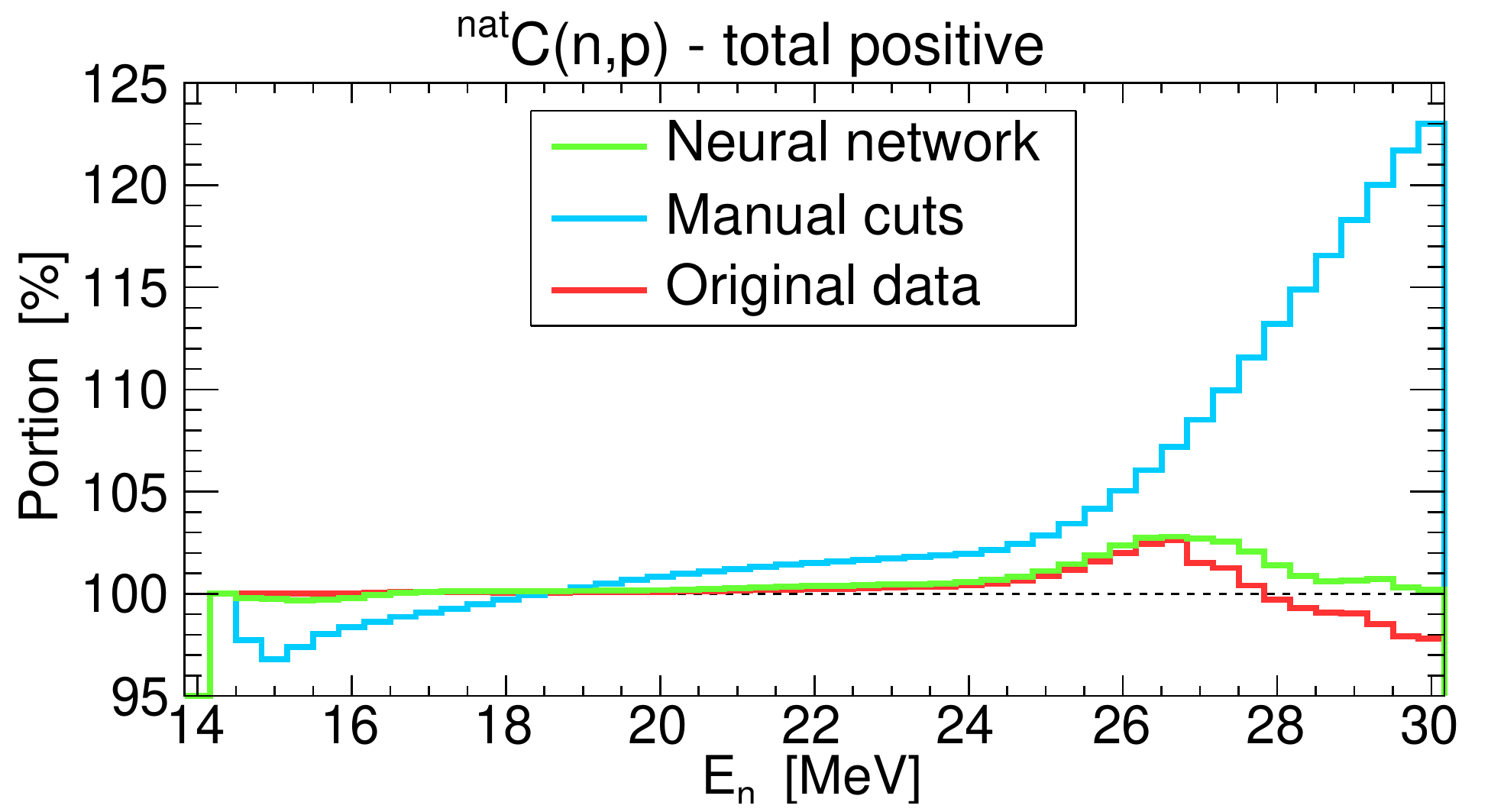}\includegraphics[width=0.39\linewidth]{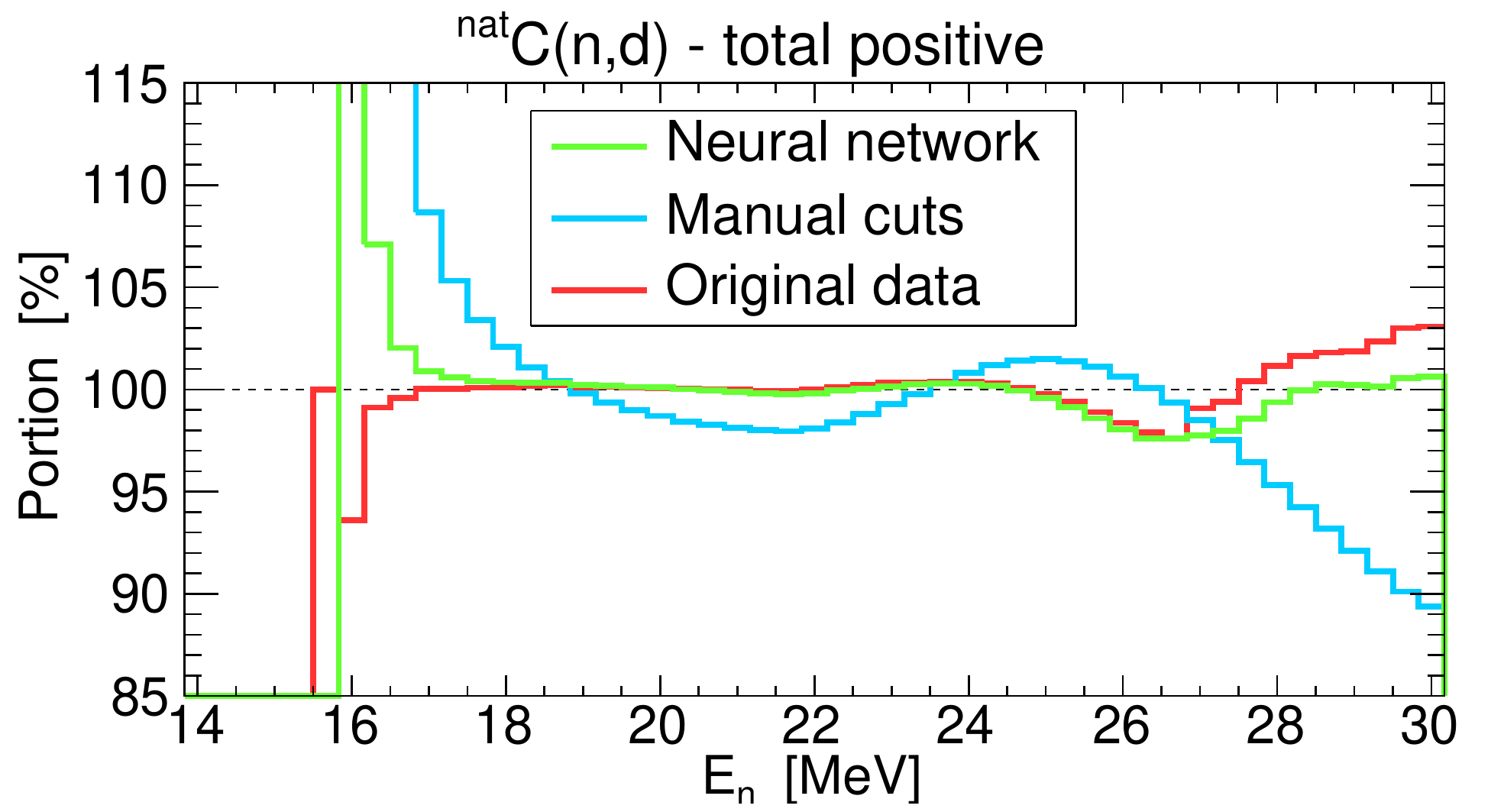}
\caption{Portions of a reaction-specific detection yield classified correctly or incorrectly, either by the trained neural networks, or by applying the manual cuts, or by the original voxel classification based on the raw detection yield data. Top row: true-positive portions $\mathrm{P}_\mathrm{T}^{(\R)}(E_n)$; middle row: false-positive portions $\mathrm{P}_\mathrm{F}^{(\R)}(E_n)$; bottom row: sum of both contributions to the positive reaction-specific classification. Left column: (\textit{n,p}) reaction; right column: (\textit{n,d}) reaction.}
\label{fig5}
\end{figure*}

%showing the overall succes, revealing either the partial cancelation of 

In order to recover a voxel classification from trained networks, we determine the voxel's type simply by checking which of the output neurons (see Fig.~\ref{fig4}) provides a maximum result. We have, of course, tested more stringent conditions such as insisting that \mbox{$\R_i>0.5>\R_j+\R_k$} for a given voxel to be classified as $i$-type (otherwise it is left undecided). However, these stricter conditions do not significantly affect the overall voxel-type recovery rate (reported below). Within a relevant portion of a parameter space these conditions only result in a few sporadic undecided voxels between the  specific reaction patterns (similar to the empty voxels between $(n,p)$ and $(n,d)$ patterns from the right plot of Fig.~\ref{fig3}). In light of no notable improvement in the classification rate, we find this behavior to work against the desired classification outcome. For this reason we adopt the simplest classification recovery scheme (by finding the maximum $\R_i$) as the most efficient one.

We now demonstrate the quality of trained neural networks and compare it with the quality of manual cuts from Figs.~\ref{fig2} and \ref{fig3}. While the convergence of network weights is satisfactory already after a couple of hundred iterations, we use in this work $10^4$ iterations\footnote{
We used the parallel processing capabilities of \texttt{TProof} class from ROOT in order to distribute the training jobs for separate $\Delta E$-$E$ pairs of strips between available CPU cores. For training we have used a table computer based on Intel Xeon E5-1603 v3(2.80~GHz) processor, which (for the reported network structure of $10\times10$ neurons) takes approximately $2\times10^{-7}$ minutes per voxel and per training iteration. For the reported number of submitted voxels (between \mbox{29\,000} and \mbox{47\,500} per $\Delta E$-$E$ pair) it takes between 55 and 90 minutes for $10^4$ training iterations to be completed for a given pair of $\Delta E$-$E$ strips.
}. For this demonstration we consider the portions of a detection yield that are correctly (true positive) or incorrectly (false positive) classified as \mbox{$\R=(n,p)$}  or \mbox{$\R=(n,d)$}  type. We first define the true-positive contributions $\mathrm{T}_\R^{(\pair)}$ to a particular reaction type:
\begin{linenomath}\begin{equation}
\mathrm{T}_\R^{(\pair)}(\Delta E,E,E_n)=\left\{\begin{array}{cl}
x_\R^{(\pair)}(\Delta E,E,E_n)&\text{if classified as } \R\\
0&\text{otherwise}
\end{array}\right. ,
\end{equation}\end{linenomath}
together with the false-positive contributions $\mathrm{F}_\R^{(\pair)}$:
\begin{linenomath}\begin{equation}
\mathrm{F}_\R^{(\pair)}(\Delta E,E,E_n)=\left\{\begin{array}{cl}
\sum_{\R'\neq R}x_{\R'}^{(\pair)}(\Delta E,E,E_n)&\text{if classified as } \R\\
0&\text{otherwise}
\end{array}\right. .
\end{equation}\end{linenomath}
We further define the correctly and incorrectly classified portions $\mathrm{P}_\mathrm{T}^{(\R)}$ and $\mathrm{P}_\mathrm{F}^{(\R)}$ of a reaction-specific detection yield:
\begin{linenomath}\begin{align}
&\mathrm{P}_\mathrm{T}^{(\R)}(E_n)=\frac{\sum_{\pair}\sum_{\Delta E,E}\mathrm{T}_\R^{(\pair)}(\Delta E,E,E_n)}{\sum_{\pair}\sum_{\Delta E,E}x_\R^{(\pair)}(\Delta E,E,E_n)},\\
&\mathrm{P}_\mathrm{F}^{(\R)}(E_n)=\frac{\sum_{\pair}\sum_{\Delta E,E}\mathrm{F}_\R^{(\pair)}(\Delta E,E,E_n)}{\sum_{\pair}\sum_{\Delta E,E}x_\R^{(\pair)}(\Delta E,E,E_n)}.
\end{align}\end{linenomath}
Though these portions may be defined for each $\Delta E$-$E$ pair $\pair$ of silicon strips separately, for demonstration purposes we have defined them as averaged over 72 relevant pairs of strips.

Figure~\ref{fig5} shows thus defined portions of true-positive and false-positive classifications. Left column and right column of plots correspond to the (\textit{n,p}) and (\textit{n,d}) reactions, respectively. Top row shows the true-positive portions $\mathrm{P}_\mathrm{T}^{(n,p)}$ and $\mathrm{P}_\mathrm{T}^{(n,d)}$. Middle row shows the false-positive portions $\mathrm{P}_\mathrm{F}^{(n,p)}$ and $\mathrm{P}_\mathrm{F}^{(n,d)}$. Bottom row shows the total amount of positive classifications, i.e. the sum of true-positive and false-positive classifications for both types of reactions. The most important comparison is that between the trained neural networks and the laboriously selected manual cuts. A quality of these two classification schemes is also compared with the initial classification based on the raw detection yield data (an example from the left plot of Fig.~\ref{fig3}), which we refer to as the \textit{original classification}. The original classification shows how much of a reaction-specific detection yield could be successfully recovered if every single voxel could be correctly recognized (e.g. by overfitted neural networks), thus serving as a reference point for evaluating the quality of any other type of classification. It should be noted that one does not wish to use the original classification for the analysis of experimental data, because it does not lend itself to any kind of generalization outside the initially identified patterns. In other words, it cannot resolve the type of any (experimentally obtained) count which falls within originally undecided voxels.

It can be readily appreciated that the classification quality of trained neural networks barely deviates from the quality of original classification. A deviation of the original classification from 100\% in true positives and from 0\% in false positives is due to the fact that within some voxels there might have been counts from multiple reactions -- e.g. a mixture of (\textit{n,p}) and (\textit{n,d}) counts -- while each voxel is assigned only a single reaction type at the end of the classification procedure. It should be noted that the level of a reaction type recovery from original classification depends on the adopted voxel size, since the overlap of counts decreases by decreasing the voxel size. Therefore, one needs to ensure that the adopted voxel density is not too dissimilar from the one used in the analysis of experimental data.

In general, one aims for as high as possible portion of true-positive classifications, and as low as possible portion of false-positive classifications. The fact that at least up to 25~MeV both the original classification and the neural network classification keep close to 100\% in true positives (top row) and to 0\% in false positives (middle row), suggests that the experimental data from a joint $^\car$C(\textit{n,p}) and $^\car$C(\textit{n,d}) measurement can be reliably analyzed (at least) up to 25~MeV. Based on these results, one can expect an uncertainty in the identification within a few percent when using the optimized neural networks.

\pagebreak

The results from trained neural networks are clearly superior to those from manual cuts, at least up to the relevant limit of 25~MeV. Even when manual cuts seem to outperform trained networks, which appears to be the case for true-positive (\textit{n,p}) classifications, they do so at the price of performing poorly at the complementary task, that of minimizing false-positive classifications. This is clearly seen from the total portion of positive classifications (bottom row) for both types of reactions. The desired outcome in these plots is to be as close to 100\% as possible. Trained networks clearly outperform manual cuts in the overall positive classification.

%The efficacy of the neural networks is not only comparable to, but in places even better than the efficacy from Fig.~\ref{fig4}. The possibility of such outcome can be easily understood from the following extreme example: if we had such an inadequate classification scheme that all the counts were classified as the same type (e.g. all counts were identified as protons), then the true positive rate for this type would be 100\%. This clearly illustrates that the adequate classification scheme may indeed outperform the initial voxel categorization, in a sense of a true positive identification.

Total portions of positive classification close to 100\% also suggest an accidental -- not to be relied upon -- but possibly favorable type of outcome to be expected in the analysis of experimental data. When one starts losing portions of the correctly classified yield due to a decrease in true-positive classifications, a \textit{simultaneous} increase in false-positive classifications may lead to a partial cancellation of these opposing effects and at least a partial restoration of a reaction-specific detection yield. For example, a portion of deuterons misclassified as protons may partially recover the portion of protons misclassified as deuterons, and vice versa. While each departure of true-positive classifications from 100\% and of false-positive classifications from 0\% is a negative effect in itself, their combined effect opens a possibility for a more optimistic outcome, conditional on the identification of the optimal type-separation boundaries.

\section{Conclusions}
\label{conclusions}

We have developed a machine learning based procedure for classification of the proton and deuteron counts from a joint $^\car$C(\textit{n,p}) and $^\car$C(\textit{n,d}) measurement from n\_TOF. An important part of the procedure is a careful preparation of training datasets. Training data, consisting of triples ($\Delta E,E,E_n$) of relevant parameters identifying each count, were obtained by Geant4 simulations of eight relevant neutron induced reactions on natural carbon: (\textit{n,p}), (\textit{n,np}), (\textit{n,d}) and (\textit{n,t}) reactions on both $^{12}$C and $^{13}$C isotopes. In order to solve practical difficulties in applying a neural network training procedure to a set of raw counts, we have first constructed a 3-parameter space and divided it into discrete voxels. Each voxel was assigned a specific type, discriminating spectral patterns of the (\textit{n,p}) and (\textit{n,d}) reactions from competing contributions of the (\textit{n,np}) and (\textit{n,t}) reactions. These spectral patterns were further refined based on careful considerations and thus obtained sets of voxels were submitted to a neural network training procedure. Spectral patterns for each relevant $\Delta E$-$E$ pair of silicon strips, 72 in total, were treated separately, each being afforded its own neural network. To this end, a \texttt{TMultiLayerPerceptron} class from ROOT was used. Both the input data and the network parameters were varied in order to identify the optimal network configuration, as well as the optimal training procedure. A stochastic minimization with manually adjusted hyperparameters \mbox{$\tau=5$} and \mbox{$\eta=0.01$} was found to provide the best performance, regarding both the convergence rate and the processing time per training iteration. A 2-layer neural network with 10 neurons in each layer was adopted as the optimal network structure. A performance of trained networks was compared against carefully determined manual cuts between reaction types, by examining the portions of true-positive and false-positive classifications for the relevant (\textit{n,p}) and (\textit{n,d}) reactions. Trained neural networks were found to be clearly superior in quality of classification and to be a basis for a reliable analysis of experimental data up to at least 25~MeV of neutron energy.

%Based on the observed separability of the (\textit{n,p}) and (\textit{n,d}) counts, we conclude that the experimental $^\car$C(\textit{n,p}) and $^\car$C(\textit{n,d}) data from n\_TOF can be reliably analyzed from the reaction threshold -- around 14~MeV for the (\textit{n,p}) and 15~MeV for the (\textit{n,d}) reaction -- up to at least 26~MeV.

{\color{white}.}

\textbf{Acknowledgements}\\

This work was supported by the Croatian Science Foundation under Project No. 8570. Geant4 simulations were run at the Laboratory for Advanced Computing, Faculty of Science, University of Zagreb.

\appendix

\section{Extracting cross sections from Geant4}
\label{cross_section}

We describe a procedure for extracting any type of cross section from Geant4 simulations. We stress that the simulations described here are completely independent of and in all aspects different from simulations described in Section~\ref{simulations}; they are in no manner related to or constrained by any particular experimental setup. We run these simulations separately for pure $^{12}$C sample and for pure $^{13}$C sample, which is of arbitrary but preferably regular shape and dimensions. For a given reaction type $\R$ on a pure isotope $\Z$, a first-chance reaction yield $Y_{\Z,\R}$ (without the multiple scattering effects) may be expressed as:
\begin{linenomath}\begin{equation}
Y_{\Z,\R}(E_n)=\left(1-e^{-\eta_\Z \sigma_{\Z}(E_n)}\right)\frac{\sigma_{\Z,\R}(E_n)}{\sigma_{\Z}(E_n)}=\frac{N_{\Z,\R}(E_n)}{\mathcal{N}_\Z(E_n)}.
\label{A1}
\end{equation}\end{linenomath}
Here $\eta_\Z$ is the areal density of a specific sample (in number of atoms per unit area), $\sigma_{\Z,\R}(E_n)$ is the cross section for a particular reaction of interest, and $\sigma_{\Z}(E_n)$ is the total cross section for any reaction at all. The sample is irradiated by neutrons (or any other relevant incident particles) following some predefined energy distribution. Reaction yield (for a given sample) may be reconstructed by counting the total number $\mathcal{N}_\Z(E_n)$ of neutrons generated with the energy $E_n$, and the number $N_{\Z,\R}(E_n)$ of first-chance reactions -- first occurrences of a specific reaction, which are counted only if that reaction was the first one to take place among all possible neutron interactions, including the elastic scattering. By first counting the number $N_{\Z}(E_n)$ all possible first-chance reactions, governed by the total cross section $\sigma_{\Z}(E_n)$, one obtains upon inverting Eq.~(\ref{A1}):
\begin{linenomath}\begin{equation}
\sigma_\Z(E_n)=-\frac{1}{\eta_\Z}\ln\left(1-\frac{N_\Z(E_n)}{\mathcal{N}_\Z(E_n)}\right).
\end{equation}\end{linenomath}
From thus reconstructed total cross section one only needs to make the scaling between specific reactions of interest and all induced reactions:
\begin{linenomath}\begin{equation}
\sigma_{\Z,\R}(E_n)=\frac{N_{\Z,\R}(E_n)}{N_\Z(E_n)}\sigma_\Z(E_n)
\end{equation}\end{linenomath}
in order to reconstruct the cross section of a particular reaction.

\end{document}